\documentclass{aa}  

\usepackage{graphicx}
\usepackage{subfigure}
\usepackage{adjustbox} 
\usepackage{txfonts}
\usepackage{amsmath}
\usepackage{txfonts}
\usepackage{xcolor}
\usepackage{bm}
\usepackage{float}
\usepackage{svg}
\usepackage{natbib}
\usepackage{booktabs}
\usepackage{fixltx2e}
\usepackage{tikz}
\usepackage{pifont}
\usepackage{array}

\usepackage{url}
\usepackage{hyperref}
\hypersetup{
   colorlinks=true,       
   linkcolor=black,       
   citecolor=blue,        
   urlcolor=black         
}
\usepackage{comment}

\newcommand{\EA}[1]{{\color{black} #1}}
\newcommand{\rev}[1]{{\color{black} #1}}

\newcommand{\figpos}[1]{{\color{orange}#1}}

\newcommand{\HI}{\ifmmode \mathrm{\ion{H}{I}} \else \ion{H}{I} \fi}

\def\GHz{\ifmmode $\,GHz$\else \,GHz\fi}
\def\MJysr{\ifmmode \,$MJy\,sr\mo$\else \,MJy\,sr\mo\fi}
\def\microns{\ifmmode \,\mu$m$\else \,$\mu$m\fi}
\def\pc{\ifmmode \,$pc$\else \,pc\fi}
\def\mpc{\ifmmode \,$mpc$\else \,mpc\fi}
\def\au{\ifmmode \,$au$\else \,au\fi}

\def\kms{\ifmmode $\,km\,s$^{-1}\else \,km\,s$^{-1}$\fi}
\def\checkmark{\tikz\fill[scale=0.4](0,.35) -- (.25,0) -- (1,.7) -- (.25,.15) -- cycle;} 

\begin{document} 
   \title{Comparing the morphology of molecular clouds without supervision}

\author{
Pablo Richard\inst{\ref{inst1}},
Erwan Allys \inst{\ref{inst1}},
François Levrier\inst{\ref{inst1}},
Antoine Gusdorf\inst{\ref{inst1}, \ref{inst2} },
Constant Auclair\inst{\ref{inst1}}}

\institute{
 \label{inst1} Laboratoire de Physique de l’École normale supérieure, ENS, Université PSL, CNRS, Sorbonne Université, Université Paris Cité, 75005~Paris, France
\and
\label{inst2} Observatoire de Paris, Université PSL, Sorbonne Université, LERMA, CNRS UMR 8112, 75014 Paris, France
}

\date{Submitted to \aap\ on July 13, 2024 / Accepted March 4, 2025}

 
  \abstract{
  {Molecular clouds are astrophysical objects whose complex nonlinear dynamics are reflected in their complex morphological features. Many studies investigating the bridge between higher-order statistics and physical properties have highlighted the value of non-Gaussian morphological features in capturing physical information. Yet, as this bridge is usually characterized in the supervised world of simulations, transferring it to observations can be hazardous, especially when the discrepancy between simulations and observations remains unknown.}
  {In this paper, we aim to evaluate, directly from the observation data, the discriminating ability of a set of statistics.}
  {To do so, we developed a test that allowed us to compare the informative power of two sets of summary statistics for a given unlabeled dataset. Contrary to supervised approaches, this test does not require knowledge of any class label or parameter associated with the data. Instead, it evaluates and compares the degeneracy levels of the summary statistics based on a notion of statistical compatibility. We applied this test to column density maps of 14 nearby molecular clouds observed by \textit{Herschel} and iteratively compared different sets of typical summary statistics.}
  {We show that a standard Gaussian description of these clouds is highly degenerate but can be substantially improved when being estimated on the logarithm of the maps. This illustrates that low-order statistics, when properly used, remain a very powerful tool. We further show that such descriptions still exhibit a small quantity of degeneracies, some of which are lifted by the higher-order statistics provided by reduced wavelet scattering transforms. \EA{These degeneracies} quantitatively differ \EA{between observations and} state-of-the-art simulations of dense molecular cloud collapse\EA{, and they are not present for} log-fractional Brownian motion models. Finally, we show how the summary statistics identified can be cooperatively used to build a morphological distance, which is evaluated visually and gives \EA{convincing} results.}
  {}
  }


   \authorrunning{Richard et al.}

   \titlerunning{Comparing the morphology of molecular clouds without supervision}

   \maketitle
%

\section{Introduction} \label{intro}

Molecular clouds (MCs) play a key role in star formation. Their highly nonlinear dynamics tend to couple spatial scales over a wide range, creating dense filamentary structures in which clumps form, and these eventually lead to prestellar cores~\citep{McKee2007}. However, the precise role of each physical ingredient (such as turbulence, magnetic fields, and gravity, to name but a few) in these dynamics still remains to be fully understood.


\begin{figure}[htbp!]
    \centering
    \includegraphics[width=\linewidth]{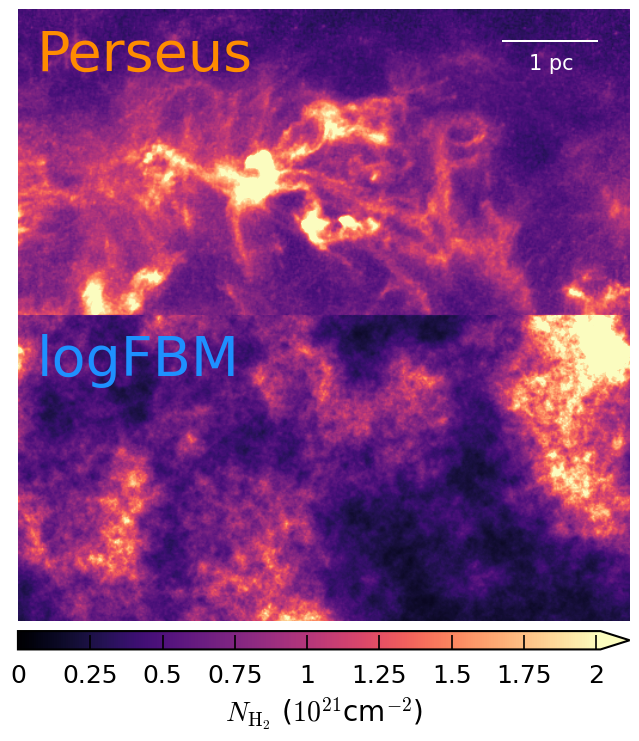}
    \caption{Column density map of a region in Perseus {(top row)} and sample of a log-fractional Brownian motion \EA{(logFBM)} field {(bottom row)}. Each image has $256\times 512$ pixels but is tiled into two complementary patches of size $256\times 256$ on which the statistics shown in Fig.~\ref{fig:ISM_is_not_logFBM_stats} are computed. While these one-point and two-point statistics (some being non-Gaussian) are clearly compatible, these two images have manifestly different morphologies.}
    \label{fig:ISM_is_not_logFBM}
\end{figure} 

\begin{figure}[htbp!]
    \centering
    \includegraphics[width=\linewidth]{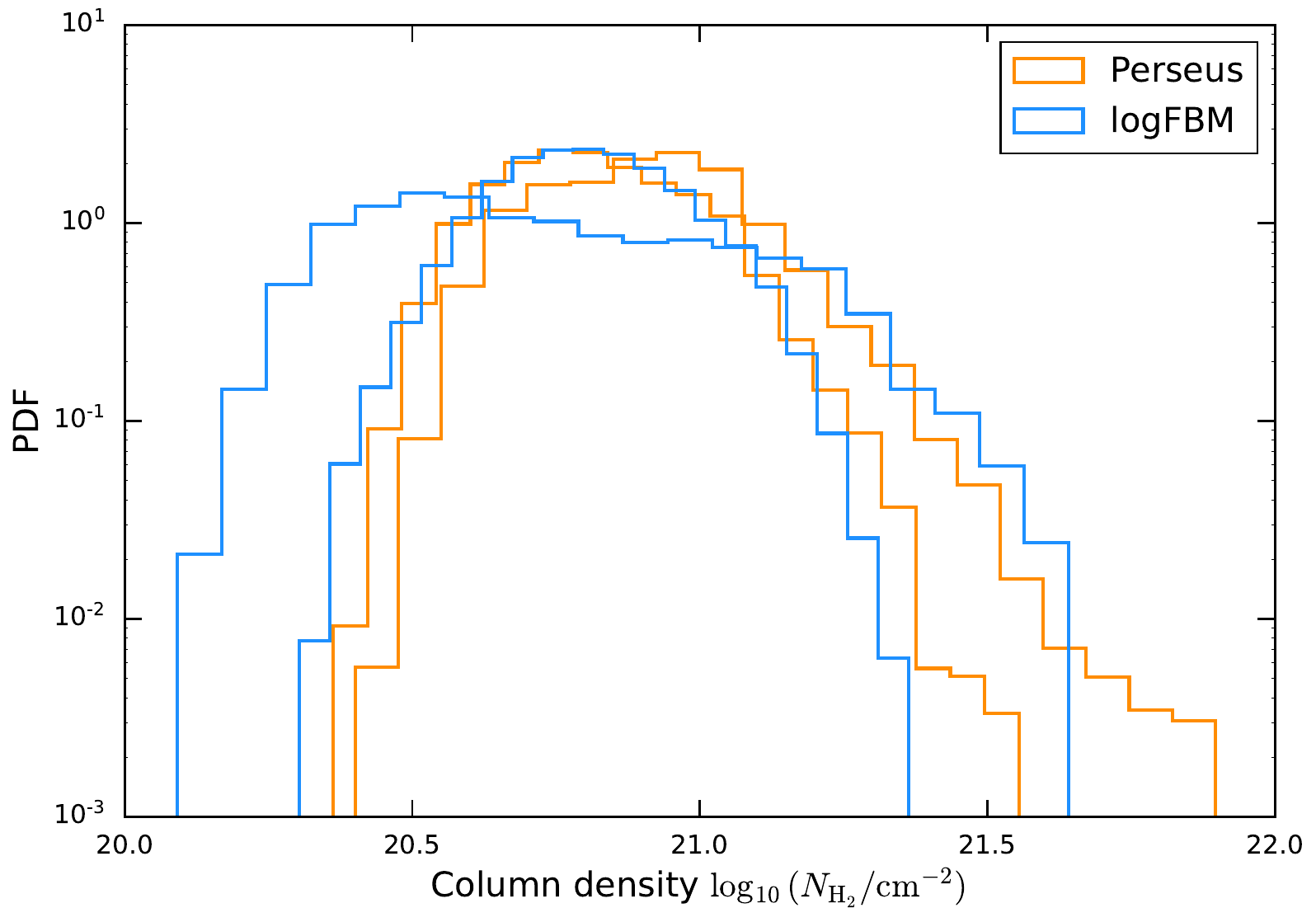}
    \includegraphics[width=\linewidth]{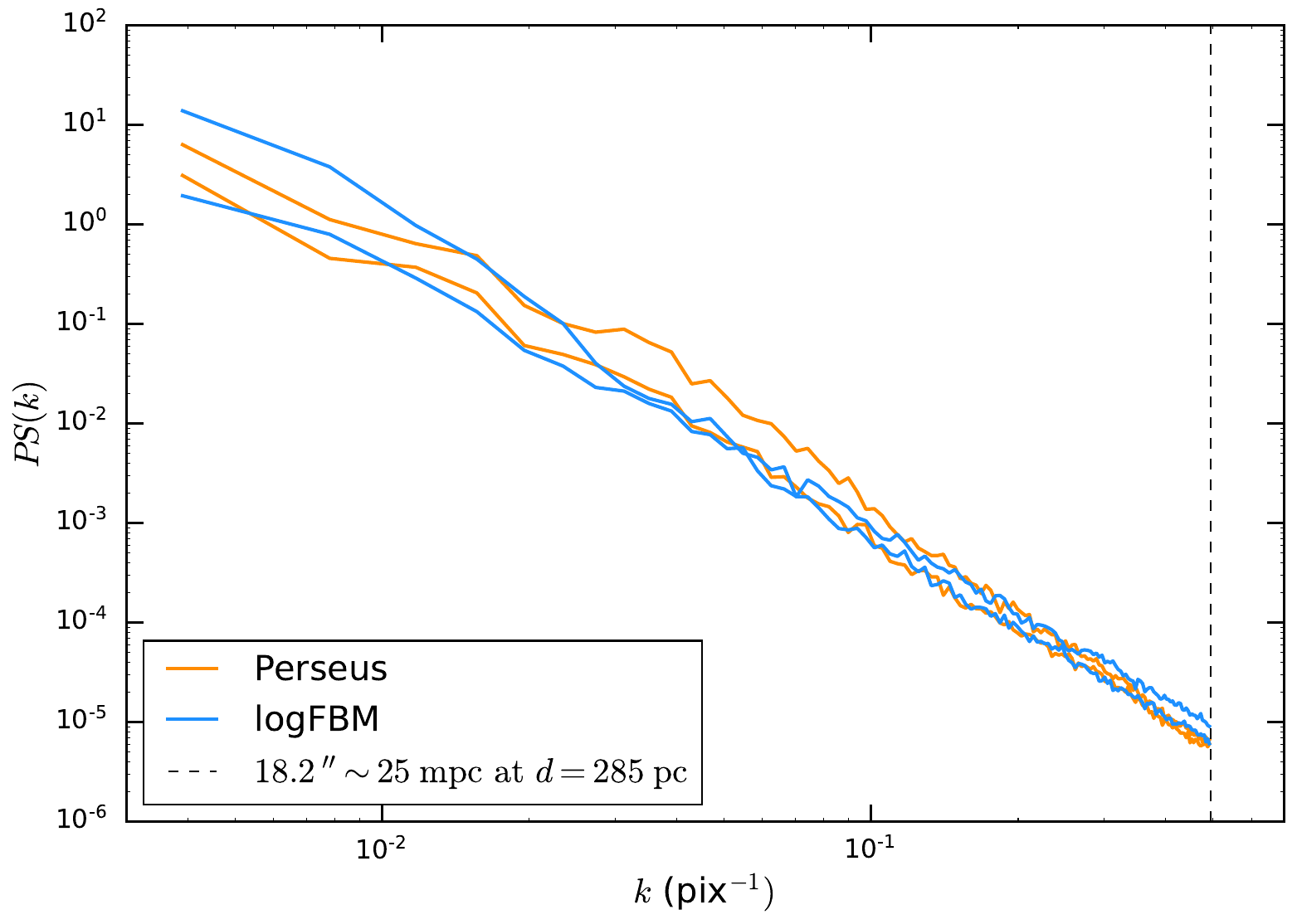}
    \includegraphics[width=\linewidth]{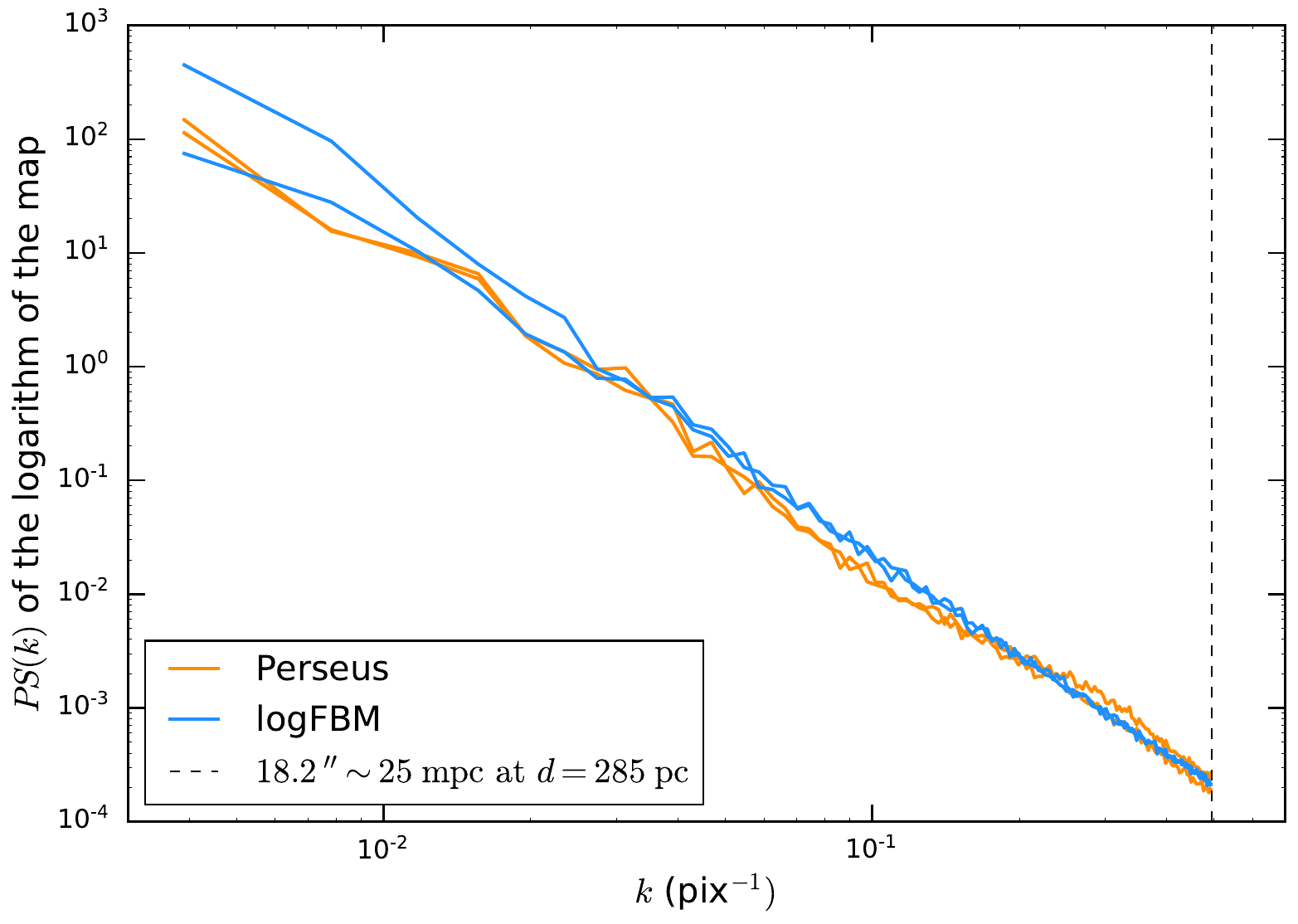}
    \caption{Statistical properties of the maps from Fig.~\ref{fig:ISM_is_not_logFBM}: PDFs {(top)}, power spectra {(center)}, and power spectra of the logarithms of the maps {(bottom)}. The dashed vertical line in the power spectra plots is located at the $18.2\arcsec$ resolution of the column density map from the observational dataset. Power spectra are apodized as explained in Sect.~\ref{sec_statistics}.
    We see that the logFBM process is compatible with this observed portion of MC for all these statistics (some of theme being non-Gaussian), while having a manifestly different morphology, as revealed in Fig.~\ref{fig:ISM_is_not_logFBM}. All these statistics are thus degenerate for such a comparison.}
    \label{fig:ISM_is_not_logFBM_stats}
\end{figure} 

A key step toward such an understanding certainly resides in the ability to decipher, in the morphology of these structured clouds, some signature of the physical processes at play.
For instance, the power-law shape of the power spectrum (PS) in emission maps may trace properties of the turbulence~(\cite{miville2007statistical, federrath2013star}), or the shape of the probability distribution function (PDF) of column density maps~\citep{vazquez1997search, hennebelle2008analytical, kainulainen2009probing, federrath2013star, schneider2022understanding, appel2022effects} may trace the impact of gravitational collapse or stellar feedback as it transitions from a log-normal shape to the development of a heavy tail. 

However, the strong nonlinear interplay between the physical processes at work leads to the emergence of non-Gaussian structures, such as filaments, in the interstellar medium (ISM). This has motivated numerous studies to capture physical information beyond one-point and two-point statistics. Many approaches have been investigated, including diagnostics of the phase coherence of Fourier modes \citep{levrier2006fourier, burkhart2016phase}, bispectrum \citep{burkhart2009density}, structure functions \citep{heyer2004universality}, dendograms \citep{goodman2009role}, multiscale segmentation \citep{robitaille2019exposing}, scattering transforms \citep{allys_rwst_2019, regaldo2020statistical, allys2020new, saydjari2021classification}, and neural networks \citep{peek2019androids, zavagno2023supervised}. 

If higher-order statistics represent an appealing way to refine the \EA{morphological} description of complex ISM structures, they are nevertheless subject to the following caveats:
\begin{enumerate}
    \item They should not undermine the potential of low-order statistics but rather be regarded as complementary diagnostics \citep{burkhart2016phase}. Low-order descriptors should not be underestimated; it is not because the processes under study are highly non-Gaussian that low-order statistics are only marginally informative. Yet, in numerous studies, such easy-access information is not considered (at times, it is even intentionally discarded\footnote{The mean and variance of the data may be normalized, the power spectra flattened, and histograms (nonlinearly) equalized.}) in order to emphasize the contribution of higher-order statistics. 
    \item Refining\label{caveat_2} the statistical description with higher-order terms complicates the connection between the statistical descriptions and the physical processes and associated parameters. Thus, such a connection is usually learned in the supervised world of simulations, \EA{typically via a Fisher analysis~\citep{allys2020new} or a classification task~\citep{saydjari2021classification}.} But transferring \EA{this connection} to observations can be hazardous, especially when their \EA{similarity} to simulations remains unknown~\citep{Falgarone2004, peek2019androids}\EA{. Indeed, the informativeness of a set of summary statistics can be highly dataset dependent~\citep{nunes2010optimal, blum2013comparative}}.
\end{enumerate}

In this paper, we aim to identify which sets of summary statistics are relevant to characterize the diversity of observed molecular clouds without relying on simulations or prior knowledge. In such an unsupervised framework, various observations cannot be grouped a priori in a similar class, so one often has to work in a very low data regime, which dramatically restricts the range of tools that can be used. 
This unsupervised and prior-knowledge-free framework excludes features that require significant tuning to extract information, such as neural networks. This priority shift from theoretical informativeness to actual information retrievable in this framework might reshuffle the cards regarding the various sets of summary statistics to consider.
This led us to wonder to what extent non-Gaussian statistics can actually bring meaningful contributions in a fully observation-driven and unsupervised framework.

To answer this question, we rely on a dataset of column density maps constructed from a survey of nearby molecular clouds. Our approach is to identify the amount of degeneracy that a set of summary statistics can have on this dataset, that is, finding a pair of maps in this dataset that are compatible according to this set of statistics but that actually have a different morphology. A simplified illustration of such a degeneracy is given in Figs.~\ref{fig:ISM_is_not_logFBM} and~\ref{fig:ISM_is_not_logFBM_stats}, where we compare an observed column density map with a realization of a field, called logFBM, whose logarithm is sampled from a fractional Brownian motion (FBM), which is a specific type of Gaussian process with a power-law PS. In this example, the PDFs of these maps, their power spectra, and the power spectra of their logarithms all appear compatible, even though the maps clearly have different structures. \EA{For instance, the Perseus cloud shows filamentary features that are absent in the logFBM sample.} This illustration underlines the limitation of these statistics, but only for this specific comparison, where we compare one observation to a specific known model. However, in an unsupervised framework, when working solely with unlabeled observations, we exhibit these limitations by introducing additional statistics that could be used to lift these degeneracies. 

In this work, we use such a degeneracy diagnostic on the observed molecular clouds dataset to progressively build a set of summary statistics \EA{that allows us to characterize their diversity, even in a low data regime}. We start with basic statistics, such as the PS and one-point statistics. We evidence strong limitations that we investigate and substantially mitigate by applying a logarithmic transformation to the maps beforehand. We exhibit further but moderate limitations of this improved set, leveraging higher-order statistics through the reduced wavelet scattering transform (RWST). This study, initially carried out only on the observational dataset, is then extended to logFBM data as well as to a set of magnetohydrodynamics (MHD) simulations intended to reproduce observations of dense star-forming molecular clouds. We conclude by discussing the relevance and limitations of the final set of statistics we have constructed and show how it can be used to assess the distance, in a statistical sense, between different maps of the ISM.

The paper is structured as follows:\begin{itemize}
    \item In Sect.~\ref{section data}, we present the set of MCs considered in this work and the corresponding column density maps as well as the numerical simulations and the logFBM models.
    \item We present in Sect.~\ref{sec_method} the diagnostic of statistical compatibility on which we rely and our general methodology.
    \item We present in Sect.~\ref{sec_statistics} the sets of statistics that are confronted.
    \item We apply this methodology in Sect.~\ref{section results} to confront these sets of statistics on observations, and from these results, we design an informative and low-dimensional set of summary statistics $\phi_\text{final}$.
    \item We finally define, in Sect.~\ref{sec_distance}, a \EA{morphological} distance based on $\phi_\text{final}$ that allows us to compare datasets, such as observations and simulations.\\
\end{itemize}

\section{Data} \label{section data}

In this section, we present the main dataset that we use for this study: an ensemble of $\sim$550 molecular hydrogen ($\mathrm{H}_2$) column density maps from a survey of nearby MCs. Three other datasets are used in this paper:
\begin{itemize}
    \item A set of $\sim$230 \EA{$\mathrm{H_2}$} column density maps built from magnetohydrodynamical numerical simulations of dense molecular clouds, classified into three subsets with varying values of the magnetic field, from null ("Hydro high res"), to medium ("MHD"), and high ("MHD high B"). These simulations are \EA{used in this work as an example of} state-of-the-art attempts to reproduce the physics at play in the early stages of star formation, including self-gravity and decaying turbulence\EA{. As it is often the case in studies of complex multi-physics ISM systems such as molecular clouds, these simulations focus on, and account only for, a partial variety of the physical processes at stake. For instance the set of simulations chosen here do not include any form of stellar feedback, that is nevertheless present in some regions of our observational dataset.}
    \item A set of $\sim$500 synthetic maps sampled from logFBM models, whose parameters reproduce the diversity of the observational dataset. This type of purely synthetic fields has already been extensively used to model the ISM~\citep[see, e.g.,][]{elmegreen2002fractal, brunt2002interstellar, miville2007statistical, levrier_statistics_2018}.
    \item A set of $\sim$1000 images from a large collection of everyday textures, the Describable Texture Dataset (DTD, \cite{cimpoi14describing}). We use these to emphasize the specificity of ISM fields in the context of image texture analysis.
\end{itemize}
More details about these sets are provided in Appendix~\ref{appendix:otherdatasets}. All these maps have a size of $512\times 512$ pixels.

\subsection{Observations: Column density maps from the HGBS 
} \label{obs}

\begin{figure*}
    \centering
    \hspace*{-2.2cm}\includegraphics[width=1.3\linewidth]{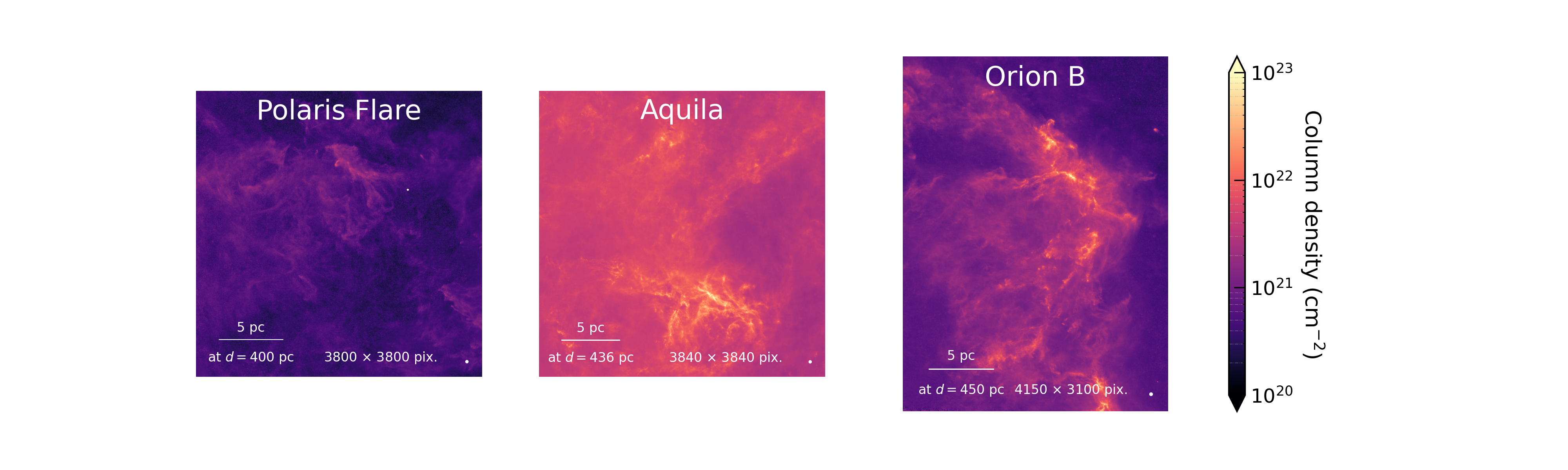}
    \caption{Three examples of \EA{$\mathrm{H_2}$} column density maps from the HGBS. The spatial resolution of the maps is $18\arcsec$, sampled with a pixel size of 3$\arcsec$. \EA{As a reference, a white dot with diameter $90\arcsec$ (which amounts to five times the resolution)} is shown at the bottom right corner of each map. Estimates of spatial scales, also shown on the maps, are based on reported distances (see Table~\ref{tab:regions_table}). The Polaris Flare is an example of a diffuse and quiescent cloud, while Aquila and Orion B are dense and very active star-forming regions.}
    \label{fig:obs_overview}
\end{figure*}

We focus our study on a set of MCs targeted by the Gould Belt Survey (HGBS) \citep{andre_polAq_2010} with the \textit{Herschel} Space Observatory \citep{Pilbratt2010A&A}\EA{, whose footprints on the sky are shown in Fig.~\ref{fig:hgbs_view}.}
The HGBS fulfills the two main criteria we require for this work:
\begin{itemize}
    \item It sampled numerous MCs with a diversity of physical and environmental conditions, from diffuse and quiescent regions with no sign of star formation activity such as the Polaris Flare \citep{Heithausen1990ApJ, andre_polAq_2010, MAMD10} to very dense and active ones such as Orion B \citep{Schneider2013ApJ} or the Aquila Rift cloud \citep{konyves_census_2015}. Examples of molecular column density maps from the HGBS are shown in Fig.~\ref{fig:obs_overview}. However, we emphasize that this survey is limited to local clouds (distances $d \leq$ 500 \pc) and does not encompass the full range of conditions expected in Galactic molecular clouds\footnote{A further study could target more distant, more massive star forming regions with interferometric observations. We postpone such a study to a future paper.}.
    \item It imaged a broad range of scales, from the full cloud size ($\sim10\,\pc$, corresponding to a few degrees for these nearby clouds) down to the $\sim0.1\,\pc$ scale of filaments \citep{arzoumanian2011characterizing, andre2014filamentary}, which is spatially resolved for the nearest clouds\footnote{The typical resolutions range from $10\arcsec$ at 70\microns\,to $36\arcsec$ at 500\microns.}, thus covering more than two orders of magnitude in spatial scales. This allows us to perform an in-depth morphological analysis, based on a local description of multiscale interactions.
\end{itemize}

\begin{table*}
    \caption{Summary of the properties of the different MCs studied in this paper, as well as their division into patches at different sampling.}
    \centering
    \begin{tabular}{cccccccccc}
    \hline
    \hline
    \rule{0pt}{3ex}
        Region & Dist. & \multicolumn{2}{c}{Coord.} & \multicolumn{5}{c}{\# patches with resolution}& References  \\ 
 & $d$ (pc)& $l$ (°)& $b$ (°)& Tot.& \small{6\arcsec}& \small{9\arcsec}& \small{12\arcsec}& \small{18\arcsec}& \rule[-1.2ex]{0pt}{0pt}\\
        \hline
        \rule{0pt}{3ex}
        Aquila & 436 & 28 &4 & 57 & 36 & 16 & 4 &1&  1\\
        Chamaeleon I & 210 & 296 &-16 & 28 & 20 & 6 & 2 &0& 2 \\
        Chamaeleon II III & 160 & 302 &-16 & 43 & 32 & 8 & 3 &0& 2 \\
        Corona Australis & 160 & 0 &-19 & 44 & 34 & 7 & 3 &0& 2 \\
        IC 5146 & 750 & 94 &-5 & 4 & 4 & 0 & 0 &0& 2 \\
        Lupus I & 182 & 339 &17 & 17 & 12 & 4 & 1 &0& 3 \\
        Musca & 200 & 301 &-9 & 6 & 6 & 0 & 0 &0& 4 \\
        Ophiuchus & 139 & 353 &17 & 61 & 42 & 12 & 6 &1&  5\\
        Orion B & 450 & 205 &-14 & 88 & 61 & 19 & 7 &1& 2, 6 \\
        Perseus & 285 & 159 &-20 & 44 & 32 & 10 & 2 &0& 2 \\
        Pipe Nebula & 180 & 359 &6 & 23 & 18 & 5 & 0 &0& 2\\
        Polaris Flare & 400 & 123 &26 & 50 & 36 & 9 & 4 &1& 2, 6 \\
        Serpens & 436 & 31 &4 & 46 & 33 & 9 & 4 &0& 1 \\
 Taurus & 145 & 170 & -16 & 40 & 30 & 6 & 4 & 0& 7 \rule[-1.2ex]{0pt}{0pt}\\
 \hline
    \end{tabular}
    \tablefoot{References in the rightmost column are for the distance estimates\EA{. These can vary by several tens of percentage depending on the line of sight. Such variability along with }uncertainties are given in the references quoted.}
    \tablebib{(1)~\cite{OrtizLeon2018}; (2) \cite{Zucker2020}; (3) \cite{Galli2013}; (4) \cite{KnudeHog1998}; (5) \cite{Mamajek2008}; (6) \cite{Schlafly2014}; (7) \cite{Yan2019}.}
    
    \label{tab:regions_table}
\end{table*}

We consider the high resolution (18\arcsec) \EA{$\mathrm{H_2}$} column density maps of 14 regions, produced with the procedure described in appendix~A of~\cite{palmeirim2013herschel} and publicly available from the \textit{Herschel} Gould Belt Survey Archive\footnote{\url{http://www.herschel.fr/cea/gouldbelt/en/Phocea/Vie\_des\_labos/Ast/\\ast\_visu.php?id\_ast=66}}.
This 18$\arcsec$ angular resolution corresponds to a $12 \mpc$ spatial resolution for the nearest clouds, such as Ophiuchus and Taurus ($d\sim 145 \pc$), and up to $40 \mpc$ for the most distant clouds, such as Orion B ($d\sim 450 \pc$). The main properties of these clouds are given in Table~\ref{tab:regions_table}. \EA{At small scales and low density, these observations are contaminated by the Cosmic Infrared Background (CIB) and the noise. This partly impedes the intercomparison of diffuse observations and their comparison with logFBM models. However, these contaminations are negligible for dense regions, which are the ones targeted by the set of simulations considered in this study.}

\subsection{Subsampling and tiling}
\label{secPreProcessing}

\begin{figure*}
    \centering
    \includegraphics[width=\linewidth]{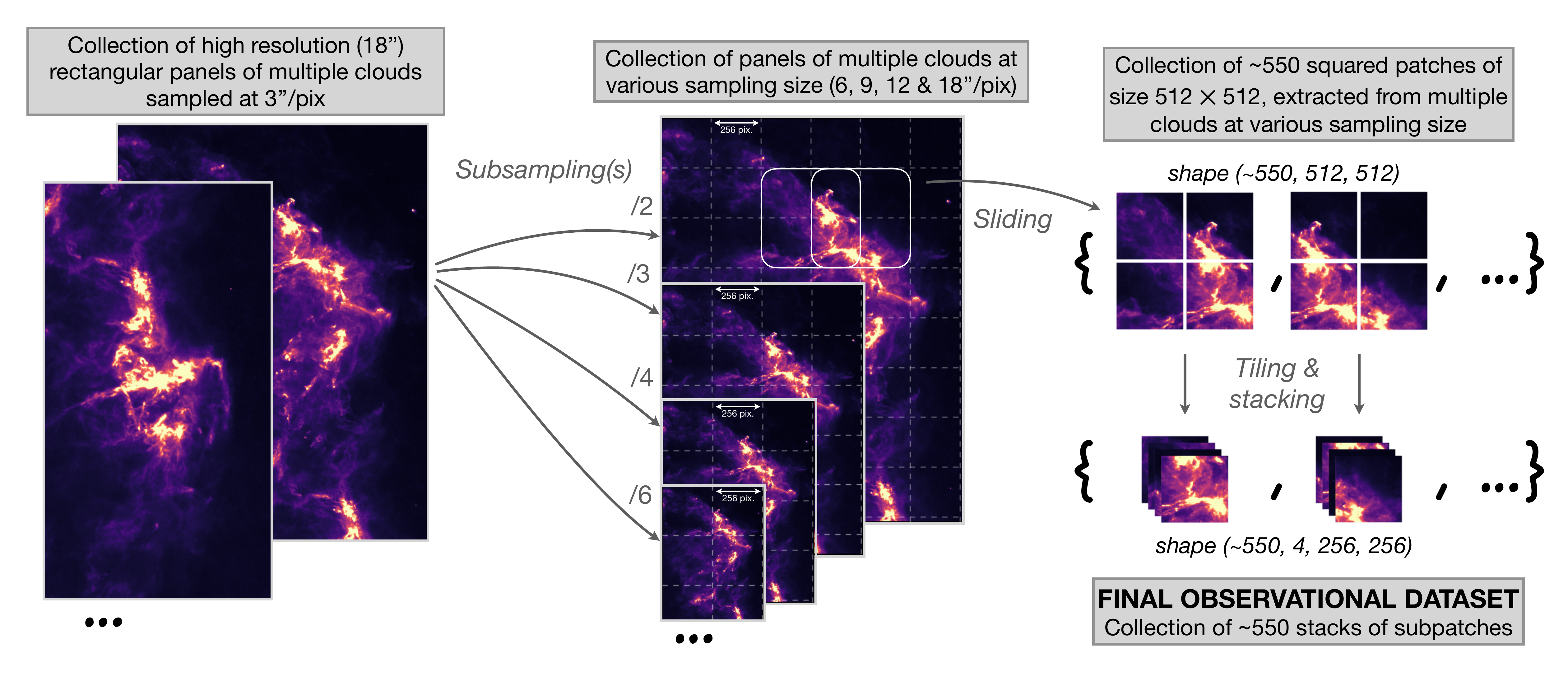}
    \caption{Illustration of the preprocessing of observational data. Details are given in Sect.~\ref{secPreProcessing}.}
    \label{fig:obs_preprocessing}
\end{figure*} 

The presence of physical processes such as gravity in the dynamics of MCs implies spatial variations of their statistical properties, which prevents us from modeling them as stationary stochastic processes.
For instance, they often exhibit strong local overdensities, while being surrounded by diffuse borders. This inhomogeneity makes it difficult to compare such objects as a whole, as well as to properly estimate their statistical properties, when seen as realizations of a random process, as these properties will, for instance, depend heavily on the identification of the cloud boundary, or on the precise definition of what a cloud is. Furthermore, variance estimates for random processes generally rely on a homogeneity assumption to estimate the intrinsic variability of a process from its spatial variations. More broadly, the estimation of the statistical moments and properties of a process is usually based on spatial averaging, which assumes the homogeneity of the sample.

To avoid this problem, we choose to restrict the comparison to local patches of these MCs, assuming statistical homogeneity within each individual patch. This requires the identification of a characteristic stationarity length over which it can be assumed that the statistical properties of the cloud do not vary significantly. Different patches extracted from the same cloud may have different statistical properties, which will be representative of the inhomogeneity of the cloud as a whole. This encourages us to use small patches. However, an additional difficulty is that these patches must be large enough to make statistical estimates. In particular, variance estimates require sampling beyond the correlation length of the process under study. 

In this paper, we have chosen to cut patches with angular sizes ranging from $0.85$° to $2.55$°, which correspond respectively to $3 \pc$ and $9 \pc$ for a cloud at a typical distance of $200 \pc$.
We believe this represents a good compromise with regard to the trade-off mentioned above. However, we are aware that there may not be an ideal solution, especially as we have no precise estimate of either the stationarity length or the correlation length. This means that we cannot rule out potential confusion between local intrinsic variability and large-scale variations in the statistical properties of the clouds. However, we believe that, for lack of a better solution, this does not undermine the relevance of this study.

More precisely, we make multiple versions of the original (18\arcsec~resolution, 3\arcsec~pixel size) column density maps at different pixel sizes (6\arcsec, 9\arcsec, 12\arcsec, and 18\arcsec). The subsampling is done using a bivariate spline approximation of order three of the original maps. Then, we cut from these maps patches of size in pixels $512\times 512$, with a step size of 256 pixels, such that two neighboring patches have 50\% of their pixels in common. In the following, these patches are tiled into four subpatches of size $256\times 256$ to perform a variability estimation. The choice of 6\arcsec~for the finest sampling allows us to exploit the 18\arcsec~resolution of these maps with mitigated sampling artifacts. We note furthermore that the smallest scales that we probe with the two-point-based statistics presented in Sect.~\ref{subsec_two_point_stats} remain above 24\arcsec. \rev{This prevents the finite 18\arcsec~resolution to significantly affect the results}. The 9\arcsec, 12\arcsec, and 18\arcsec~pixel sizes respectively correspond to a relative shrinking of the 6\arcsec~pixel maps with ratios of 1.5, 2, and 3, which allows us to accommodate for the significant and quite uncertain range of distances of the different MCs in the HGBS. This entire procedure, which is illustrated in Fig.~\ref{fig:obs_preprocessing}, and summarized in Table~\ref{tab:regions_table}, leads to a total of $551$ patches with size $512\times 512$ pixels, each of which is then subdivided into four $256\times 256$ subpatches.

\section{Quantifying informative power of summary statistics on an unlabeled dataset} \label{sec_method}

\subsection{General methodology}

\EA{We want to quantify and compare the informative power of different sets of statistics on a given dataset, that is, their ability to distinguish between patches of different physical properties.} However, because we work with unlabeled data, we cannot use supervised frameworks, such as Fisher analysis, that have a label-based approach to quantify information. For instance, when working with simulations, each data sample ${x_i}$ can be labeled by its corresponding physical parameters $\theta_i$. In our case, we have to deal with an ensemble of unlabeled maps, for which we have a priori no notion of \EA{similarity} between pairs.

In this paper, we choose to rely on a notion of compatibility between patches, \EA{computed from their summary statistics, and} that can be estimated for any set of statistics. This approach allows us to be quantitative, even in \EA{such} an unsupervised setting. Nevertheless, without supervision, it remains difficult to interpret a compatibility result between two patches for a given set of summary statistics. Indeed, it is unclear whether the two patches actually have similar properties or if they possess genuinely distinct properties whose differences are not effectively captured by the statistics.

\EA{To overcome this difficulty, we compare the results obtained from complementary sets of statistics. Indeed, we assume that if two patches are distinguished by a first set of statistics, this is sufficient to assess that these patches have different physical properties}\footnote{\EA{Indeed, this implication, from statistical to physical incompatibility, is ensured for patches that undergo stationary dynamics, both in} \EA{space and time. However, for dense molecular clouds, this assumption is not entirely satisfied when nonstationary processes such as self-gravity come into play. A potential caveat could then be to find incompatibilities between summary statistics which are not particularly relevant from a physical point of view. While we do not further discuss this issue in the present paper, we think it could deserve further work.}} and therefore to highlight the degeneracies of another set of statistics. It is this comparative approach, which is all the more relevant when the panel of statistics compared is comprehensive, that we use in this paper.

\EA{We also emphasize that} the informative power of a set of statistics strongly depends on the family of processes studied. For instance, we know that the empirical mean and PS are sufficient statistics for stationary Gaussian fields~\citep{thomas2006elements}, which is not the case for physical processes in general. In this paper, we apply our approach to the dataset of observations defined above as well as to numerical simulations and synthetic logFBM models. In addition, we note that this diagnostic can also be used to compare maps from two different datasets.

In the rest of this section, we introduce the compatibility criterion, and explain how we extend it from a \EA{level of pair of patches} to a dataset level. The sets of statistics used in this paper are presented in the following section.

\subsection{Statistical compatibility for a pair of patches}

We aim to measure a notion of compatibility, according to a given set of summary statistics $\phi$, between the two processes that generated the patches $(x_i, x_j)$. To estimate this $\phi$-compatibility, we need to make a number of simplifications, given the low data regime, that will bring us back to a simplified case of statistical hypothesis testing. To do so, we tiled each patch $x_i$ into four subpatches $\{x_i^{(l)}\}_{1\leq l \leq 4}$, as illustrated in Fig.~\ref{fig:obs_preprocessing}. We then computed the statistics at this subpatch level, $\{\phi(x_i^{(l)})\}_{1\leq l \leq 4}$, and we assumed that these random variables can be considered as independent samples of the same distribution, which we furthermore modeled as a multivariate normal distribution\footnote{This assumption is not far from being true when the law of large numbers can be applied to the distribution of $\phi(x)$. This is for instance the case when $\phi(x)$ is defined as an average over the image pixels of a certain local distortion (such as filtering, possibly nonlinear) $\phi_{loc}(x)$ of $x$: $\phi(x)=\langle \phi_{loc}(x)[\vec u]\rangle_{\vec u}$ and when the process $\phi_{loc}(x)$ is stationary with a correlation length that is small compared to the image length.} of mean $\mu_i$ and variance $\Sigma_i$:
\begin{equation}
\label{EqGaussianDistStats}
    \phi(x_i^{(l)}) \sim \mathcal{N} (\mu_i, \Sigma_i).
\end{equation}

Under these assumptions, the problem \EA{of estimating the compatibility between the distributions of $\phi(x_i^{(l)})$ and $\phi(x_j^{(l)})$} boils down to testing the compatibility between the two normal distributions, that is, to test the hypothesis: $\mu_i=\mu_j$ and $\Sigma_i=\Sigma_j$. In our case, however, we have to estimate this compatibility from very few samples, most of the time fewer than the dimension of the vector of statistics $\phi$. Thus, we choose to focus only on testing if the means $\mu_i$ and $\mu_j$ are statistically compatible, but not $\Sigma_i$ and $\Sigma_j$, a problem known as the multivariate two-sample mean test. The most widely used test statistic for this problem is Hotelling's two-sample $T^2$-statistic, a multivariate extension of Student’s $t$-test~\citep{hotelling1931generalization}. However, this test statistic requires to invert an estimation $S$ of the full covariance matrix $\Sigma_i+\Sigma_j$, which is usually intractable in our low data regime. 

To overcome this, \cite{srivastava2008test} proposed a test statistic based only on the diagonal $D_S$ of the covariance estimator $S$ \EA{(defined in Appendix~\ref{App_TestS&D})}, and on the trace of the square of its associated correlation matrix $R\equiv D_S^{-1/2}SD_S^{-1/2}$:
\begin{equation}\label{test_stat}
    d_\phi^2(x_i, x_j) \equiv \alpha \Big[\big( \hat\mu_i-\hat \mu_j \big)^T D_S^{-1}  \big( \hat\mu_i-\hat\mu_j \big)-\beta\Big],
\end{equation}
where $\hat{\mu}_i$ is the estimator of $\mu_i$ obtained through an average over the four subpatches $x_i^{(l)}$, $\alpha$ is an overall factor to normalize the variance of $d^2_\phi$, and $\beta$ is a debiasing term. We emphasize that, through a dependence on $\operatorname{tr} (R^2)$, the $\alpha$ factor accounts, at least partially, for the correlation structure of $\phi$. Further details about these terms are provided in Appendix~\ref{App_TestS&D}.

Under the assumption that $\Sigma_i=\Sigma_j$, the $d_\phi^2(x_i, x_j)$ test statistic has a variance of order unity. Thus, when it is much larger than one, $\mu_i$ and $\mu_j$ cannot be considered compatible:
\begin{equation}
    d_\phi^2(x_i, x_j) \gg 1 \implies \mu_i  \text{ and } \mu_j \text{ are incompatible};
\end{equation}
whereas when it is of the order of one or less, it is not possible to detect a discrepancy between the two means $\mu_i$ and $\mu_j$ with the available amount of data:
\begin{equation}
\begin{aligned}
    d_\phi^2(x_i, x_j) \lessapprox 1 \implies \mu_i \text{ and } \mu_j \text{ are not incompatible based on} \\\text{the available amount of data.} 
\end{aligned}
\end{equation} 

\subsection{Comparing summary statistics on a dataset}\label{SubSec_methodo_comparing}

\begin{figure*}
    \centering
     \includegraphics[width=\linewidth]{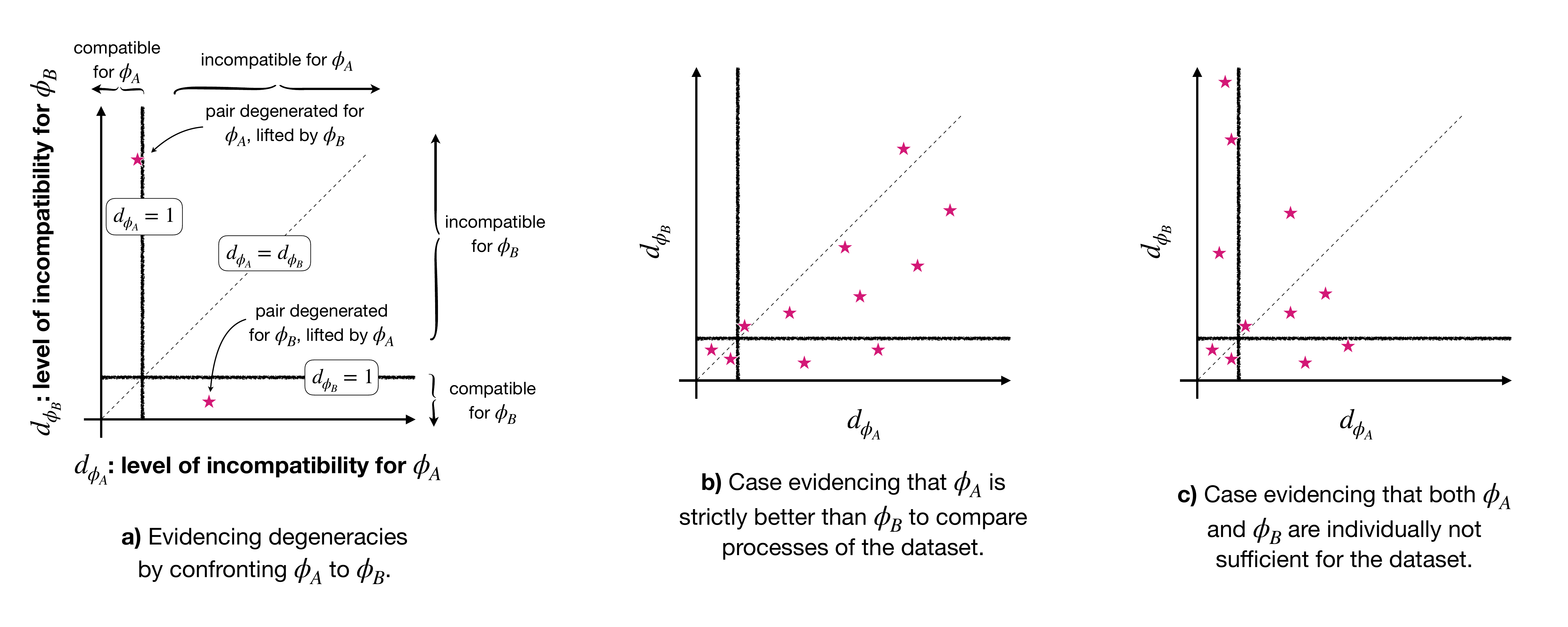}
    \caption{Illustration of the proposed test to confront two sets of summary statistics, $\phi_A$ {versus} $\phi_B$, on their degeneracy level for a given dataset. Each star represents a pair of patches. Panel~a: the presence of stars in the bottom right region, i.e., where $d_{\phi_A} \gg 1$ and $d_{\phi_B} \lessapprox 1$, reveals that some pairs of this dataset are identified by $\phi_A$ as incompatible but not by $\phi_B$: such pairs thus evidence degeneracies of $\phi_B$ lifted by $\phi_A$. Conversely, the presence of stars in the top left region evidences degeneracies of $\phi_A$ lifted by $\phi_B$. Hence, if all the points land mainly in the sub-diagonal part (panel~b), this evidences that $\phi_A$ is better suited than $\phi_B$ to compare the pairs of this dataset. If on the contrary the points are spread both in the upper left and bottom right regions of the plot (panel~c), this shows that both $\phi_A$ and $\phi_B$ are individually not sufficient to describe the processes of this dataset.}
    \label{fig:illustration_test}
\end{figure*}

By extending the $\phi$-compatibility test introduced above, we set up a comparison between two sets of summary statistics $\phi_A$ and $\phi_B$ on a given dataset. This comparison consists in studying whether each set of summary statistics has degeneracies that the other set can lift. These degeneracies are evidenced by the presence in the dataset of pairs of maps that are clearly incompatible for $\phi_A$, but for which $\phi_B$ detects no incompatibility, or vice versa. For this comparison to be relevant, the dataset must contain maps with genuinely different properties, which is the case for the sets studied below.

In practice, we suggest the following algorithm:
\begin{itemize}
    \item Step 1: For every patch $x_i$ of the dataset, compute the statistics $\phi_A$ and $\phi_B$ of its corresponding subpatches $\{x_i^{(l)}\}_l$.
    \item Step 2: For every pair $\{x_i, x_j\}$ with $i\neq j$, compute $d_{\phi_A}^2(x_i, x_j)$ and $d_{\phi_B}^2(x_i, x_j)$ from the quantities derived in step 1. 
    \item Step 3: Place every pair as a point on a 2D scatter plot showing $d_{\phi_A}^2(x_i, x_j)$ against $d_{\phi_B}^2(x_i, x_j)$, such as Fig.~\ref{fig:illustration_test}. 
\end{itemize}
The resulting scatter plot can then be used to detect whether there are some pairs of processes in the dataset that are degenerate for $\phi_A$ but \EA{not for} $\phi_B$ (or vice versa), as explained in Fig.~\ref{fig:illustration_test}. In such plots, the presence of points in the bottom right region, that is, where $d_{\phi_A} \gg 1$ and $d_{\phi_B} \lessapprox 1$, reveals that some pairs of this dataset are identified by $\phi_A$ as incompatible but not by $\phi_B$: such pairs thus evidence degeneracies of $\phi_B$ lifted by $\phi_A$ (panel~a). Conversely, the presence of points in the top left region evidences degeneracies of $\phi_A$ lifted by $\phi_B$. Hence, if all the points land mainly in the subdiagonal part (panel~b), this evidences that $\phi_A$ is better suited than $\phi_B$ to compare the pairs of this dataset. If, on the contrary, the points are spread both in the upper left and bottom right regions of the plot (panel~c), this shows that both $\phi_A$ and $\phi_B$ are individually insufficient to describe the processes of this dataset.

\section{Summary statistics} \label{sec_statistics}

We present below the different summary statistics that are used in this paper. In the following, $x$ is an image, $x(\vec u)$ is the value of the image at pixel $\vec u$, $\tilde x(\vec k)$ is the discrete Fourier transform of the image $x$ evaluated at wavevector $\vec k$, $\star$ stands for the convolution operator, and $\langle \rangle_{\vec u}$ for the averaging over pixels. When $\phi(x)$ is multivariate, $\phi(x)[i]$ is the value of its $i$-th dimension. $\bar x$ designates the following normalization of $x$: $\bar x\equiv x/\text{std}(x)$ where $\text{std}(x)\equiv \sqrt{\big\langle \big[x-\langle x \rangle_{\vec u}\big]^2 \big\rangle_{\vec u}}$.

We note that the authors consider the logarithms of some typical statistics, instead of their raw values. For instance, we use $\log\, \langle x \rangle_{\vec u}$ instead of $\langle x \rangle_{\vec u}$. This is possible when we work with positive-valued statistics. 
We made this choice to better fit the Gaussianity assumption given in Eq.~\eqref{EqGaussianDistStats}. In addition, \EA{$\phi(x)$ and $\log \, \phi(x)$ should hold the same amount of information as one can be retrieved from the other}, see Appendix~\ref{AppLogStat} for more details. \EA{For readability, we name in the following "$\phi$" instead of "log of $\phi$". We note, however, that neither $ \phi(x)$ nor $\log(\phi(x))$ should be mistaken for $ \phi(\log x)$, and we name the latter "$\phi$ of log".} If no precision is made, $\log$ designates the logarithm in base ten.

\subsection{One-point based statistics}

We list below the one-point statistics that are used in this paper. Even though some of these statistics can probe non-Gaussian information such as sparsity, they are all pointwise statistics. This means in particular that they cannot capture spatial arrangement in the maps.

\begin{itemize}
    \item The mean: \begin{equation} \boxed{
    \phi_{\text{mean}}(x) \equiv \log \,\langle x \rangle_{\vec u}.}
\end{equation}
    \item The variance: \begin{equation} \boxed{
    \phi_{\text{var}}(x) \equiv \log \, \big\langle \big[x-\langle x \rangle_{\vec u}\big]^2 \big\rangle_{\vec u}.}
\end{equation}    
    \item The mean of the logarithm: \begin{equation}\boxed{
    \phi_{\text{mean of log}}(x) \equiv \langle \log x \rangle_{\vec u}.}
\end{equation}
    \item The variance of the logarithm: \begin{equation}\boxed{
    \phi_{\text{var of log}}(x) \equiv \log \, \big\langle \big[\log x-\langle \log x \rangle_{\vec u}\big]^2 \big\rangle_{\vec u}.}
\end{equation}    
    \item The quantile functions\footnote{We consider quantile functions rather than probability distribution functions, as the difficulty of defining a unique binning range makes comparisons less efficient. Indeed, to define a range in terms of quantiles allows us for a natural adaptation to each MCs' regions.} (QF) normalized by the median: \begin{equation}\boxed{
    \phi_{\text{QF}}(x)[i] \equiv \log \,\big[ q_{\alpha_i}(x)/q_{1/2}(x)\big],}
\end{equation}
where $q_\alpha(x)$ designates the $\alpha$-quantile of the distribution of values $\{x(\vec u)\}_{\vec u}$. Hence $q_{1/2}$ stands for the median operator. We consider ten quantiles $\{\alpha_i\}_i$ such that $1-\alpha_i$ are logarithmically spaced between $10^{-4}$ and $0.4$. Low quantiles are not considered, because, \EA{for diffuse observational data, they are contaminated by the noise and the CIB}~\citep{ossenkopf2016reliability}. This logarithmic binning of the high column density values is motivated by the log-normal to power-law behavior of the tail of MCs' PDFs. See for instance~\cite{Pouteau2023} and references therein. \EA{Using quantile functions also allows the description to be very robust to outlying pixels, which in practice only drive the last quantile}.
\end{itemize}

\subsection{Two-point based statistics} \label{subsec_two_point_stats}

A very popular way to describe spatial properties of a process is through the PS. It is defined as the Fourier transform of the autocorrelation function, and describes the energy distribution scale by scale in the process studied. In this paper, we consider the isotropic PS, defined as
\begin{equation*}
    PS[k]\equiv\langle \mathbb E[|\tilde x(\vec k)|^2] \rangle _{\|\vec k\|=k}.
\end{equation*}
A power-law behavior $PS[k] \propto k^{-\beta}$ is expected for fields arising in turbulent MHD~\citep{schekochihin2022mhd}. Thus we adopted a log-log representation to linearize it and for stability purposes~\citep{bruna2013invariant}:
\begin{equation}\boxed{
    \phi_{\text{PS}}(x)[i] \equiv \log \, \big\langle | x_\text{apo}\star \psi_i|^2 \big\rangle_{\vec u},}
\end{equation}
where $x_\text{apo}$ is apodized to mitigate non periodic boundary conditions (PBC) as explained in Appendix~\ref{sec_apo}. We use six band-pass filters $\{\psi_i\}_{1\leq i\leq6}$ defined as
\begin{equation}
    \tilde{\psi}_i [\vec k] = \mathrm{1}_{\|\vec{k}\|\in [k_i, k_{i+1}[},
\end{equation}
where $k_i$ is logarithmically spaced between $k_{\min}=1/256 \, \text{pix}^{-1}$ and $k_{\max}=1/4 \, \text{pix}^{-1}$. For the finest maps with pixel size $6\arcsec$, this $k_{\max}$ corresponds to a smallest angular scale of $24\arcsec$, which remains above the $18\arcsec$ resolution of the observations.

Because some MCs tend to have log-normal behavior in their one-point statistics, we also consider PS statistics of the logarithm of the maps:
\begin{equation}\boxed{
    \phi_{\text{PS of log}}(x)[i] \equiv \log \, \big\langle |(\log x)_\text{apo}\star \psi_i|^2 \big\rangle_{\vec u}.}
\end{equation}

\subsection{Scattering statistics}

The wavelet scattering transform (WST) is a set of non-Gaussian descriptors with a hierarchical and multiscale structure~\citep{bruna2013invariant}. It has been shown to be highly effective in describing astrophysical fields~\citep[see, for instance,][]{allys_rwst_2019,saydjari2021classification}. The usual WST consists in two layers of statistics, the first of which depends on a single scale of length~$\simeq 2^j$ pixels with orientation $\theta$:\begin{equation*}
    S_1(x)[j,\theta] \equiv \langle | x\star \psi_{j,\theta}|\rangle_{\vec u}.
\end{equation*}
The second layer probes a coupling between two oriented scales $(j_1,\theta_1)$, and $(j_2,\theta_2)$, with $j_1 < j_2$: 
\begin{equation*}
     S_2(x)[j_1,j_2,\theta_1, \theta_2] \equiv  \big\langle \big| | x\star \psi_{j_1,\theta_1}| \star \psi_{j_2,\theta_2} \big| \big\rangle_{\vec u} \, / \, S_1(x)[j_1,\theta_1].
\end{equation*}
Then, observing that many processes of interest in astrophysics and cosmology exhibit strong regularities in their angular dependencies,~\cite{allys_rwst_2019} proposed the reduced WST (RWST): an angular compression of WST statistics for 2D data.  We are going to use three of the main descriptors they introduced (we refer to the above reference for more details): 
\begin{itemize}
    \item $S_1^{Iso}$, a scale-by-scale isotropic descriptor, that we normalize by estimating it on $\bar{x}$ instead of $x$:
    \begin{equation}\boxed{
    S_1^{Iso}(x)[j] \equiv \big\langle \log_2 \, \langle |\bar x\star \psi_{j,\theta}|\rangle_{\vec u} \big\rangle_\theta.}
\end{equation}We note that it computes a $L^1$ norm of the filtered field, in contrast with the PS that probes a $L^2$ norm.
\item $S_2^{Iso1}$ and $S_2^{Iso2}$, that measure an isotropic coupling between scales and are \EA{obtained by a nonlinear fit of the following model~\citep{allys_rwst_2019}}:
\begin{multline}
    \log_2 \, S_2(x)[j_1,j_2,\theta_1, \theta_2] = S_2^{Iso1}(x)[j_1,j_2] \,+ \,\\ S_2^{Iso2}(x)[j_1,j_2]\cos \big[2(\theta_2 - \theta_1)\big].
\end{multline}
They characterize, respectively, a coupling between non-oriented scales and the relative coupling between parallel and perpendicular scales.
\end{itemize}

In this paper, we furthermore compress both $S_2^{Iso1}$ and $S_2^{Iso2}$ coefficients that depend on two scales $(2^{j_1},2^{j_2})$ by only keeping their $j_2-j_1$ dependency, that is, a dependency on their ratio, by considering the following average:
\begin{equation}\boxed{
    \langle S_2^{Iso1} \rangle_{j_2-j_1}(x)[\delta] \equiv \langle S_2^{Iso1}(x)[j_1,j_2]\rangle_{j_2-j_1=\delta},}
\end{equation}
and
\begin{equation}\boxed{
    \langle S_2^{Iso2} \rangle_{j_2-j_1}(x)[\delta] \equiv \langle S_2^{Iso2}(x)[j_1,j_2]\rangle_{j_2-j_1=\delta}.}
\end{equation}

In practice, we divided the half-plane $[0,\pi[$ in eight different $\theta$ angles, and we considered four scales\footnote{\EA{This choice of $j_{\min}$ is such that, for the finest maps of pixel size $6\arcsec$, the dyadic scale $j_{\min}$ corresponds to $24\arcsec$, which remains above the $18\arcsec$} \EA{resolution of the observations. Choosing a larger $j_{\max}$ would allow us to consider larger scales. However, such scales often bring limited additional information due to their large variance, while for computational purposes, a larger $j_{\max}$ leads to a significant loss of the surface used in the image $x$ to estimate the scattering coefficients in the non PBC case. Indeed, to mitigate the latter, convolutions are done with local wavelets and then corrupted edges are cropped with a margin size $\propto 2^{j_{\max}}$.}} between $j_{\min}=2$ and $j_{\max}=5$. This leads to the three possible values $1\leq \delta \leq 3$ for our reduced $S_2$ coefficients. 
These computations are performed using the {\tt pywst}\footnote{\url{https://github.com/bregaldo/pywst}} Python package~\citep{regaldo2020statistical}. 

\subsection{Overview}

The statistics introduced above are compiled in Table~\ref{tab:stat_tools}.
In the following, in addition to the sets of summary statistics defined above, we also consider some aggregated sets of summary statistics that are made from groups of these building blocks. For instance, we refer to $\phi_{\text{Gaussian}}$ statistics to mean the joint set of \EA{$\{\phi_{\text{mean}}, \phi_{\text{PS}}\}$}. These groups are defined in Table~\ref{tab:stat_groups}.

\begin{table}[h]
    \caption{Overview of the sets of summary statistics used.}
    \centering
    \begin{tabular}{ccccc}
    \hline
    \hline
    \rule{0pt}{3ex}
        $\phi$  &\small{Dim.}& Non- & Scale & Scale  \\        
          && \small{Gaussian}& \small{description}& \small{coupling} \rule[-1.2ex]{0pt}{0pt}\\
        \hline \rule{0pt}{3ex}    
 mean&1& \small\ding{53}& \small\ding{53}& \small\ding{53}\\
 var&1& \small\ding{53}& \small\ding{53}& \small\ding{53}\\
 mean of log &1& \small{log-Gauss.}& \small\ding{53}& \small\ding{53}\\
 var of log&1& \small{log-Gauss.}& \small\ding{53}& \small\ding{53}\\
        QF&10& \checkmark& \small\ding{53}& \small\ding{53}\\
        PS&6& \small\ding{53}& \checkmark& \small\ding{53}\\
PS of log&6& \small{log-Gauss.}& \checkmark& \small\ding{53}\\
        $S_1^{Iso}$ &4& \checkmark& \checkmark& \small\ding{53}\\
        $\langle S_2^{Iso1} \rangle_{j_2-j_1}$ &3& \checkmark& \small\ding{53}& \checkmark\\
 $\langle S_2^{Iso2} \rangle_{j_2-j_1}$ &3& \checkmark& \small\ding{53}& \checkmark \rule[-1.2ex]{0pt}{0pt}\\ 
 \hline
    \end{tabular}
    \\
    \tablefoot{We report their dimension, whether they are sensitive to non-Gaussianity, provide a multiscale description, and probe couplings between scales. \EA{Here, “log-Gauss.” reminds us that, even if the statistics at stake are not Gaussian, they remain a very partial diagnostic of non-Gaussianity as these are Gaussian with respect to the logarithm of the field.}}
    \label{tab:stat_tools}
\end{table}

\begin{table}[h]
    \caption{Aggregated sets of summary statistics used in this work.}
    \centering
    \begin{tabular}{c>{\centering\arraybackslash}p{0.55\linewidth}c}
        \hline        
        \hline
        \rule{0pt}{3ex}
        Set of stats.& Composed of&Dim.\\
        \hline
        \rule{0pt}{3ex}
 Gaussian& mean + PS&7\\
 log-Gaussian& mean of log + PS of log&7\\
 RWST& $S_1^{Iso}$ + $\langle S_2^{Iso1} \rangle_{j_2-j_1}$ + $\langle S_2^{Iso2}\rangle_{j_2-j_1}$&10\\
 final& log-Gaussian + RWST&17 \rule[-1.2ex]{0pt}{0pt}\\
        \hline
    \end{tabular}
    \label{tab:stat_groups}
\end{table}

\section{Toward a low-degeneracy set of statistics}\label{section results}

We wanted to construct a set of informative summary statistics for the observational dataset. To do so, we followed a bottom-up approach, starting with the typical low-order statistics and increasingly trying to improve on them by exhibiting and lifting potential degeneracies. The underlying idea is that low-order statistics are good candidates to concentrate most of the informative power into a few coefficients. Such a concentration property is of interest, especially as our low data regime prevents us from learning many features. Another benefit of this approach is to first promote simple statistics, and use more elaborate ones only if they add a significant contribution.





\subsection{Molecular clouds have Gaussian degeneracies}

\begin{figure}[htbp!]
\includegraphics[width=\linewidth]{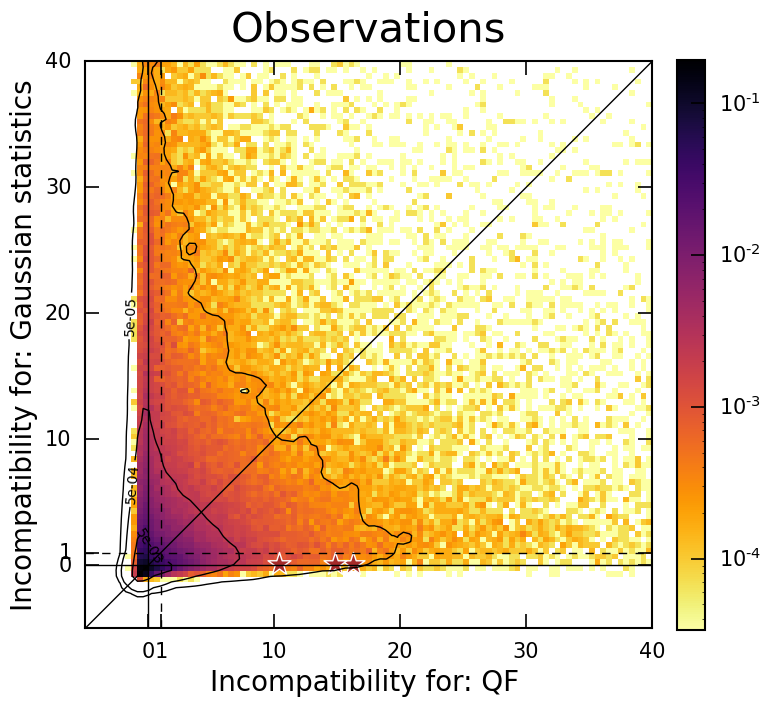}
\caption{Confronting Gaussian statistics with QF statistics on observational data, based on the test presented in Fig.~\ref{fig:illustration_test}. Each set of statistics has strong degeneracies lifted by the other set. To investigate the Gaussian confusions, we pick \EA{three} pairs of $512\times 512$ patches, whose locations on the scatter plot are given by the red stars. These pairs are shown in Fig.~\ref{fig:gaussStat_vs_pdf-1}.}
\label{fig:gaussStat_vs_pdf}
\end{figure}

To begin with, we investigate whether Gaussian statistics (i.e., the set made of mean and PS statistics) are degenerate for the observational data. To point out such potential degeneracies, we confront these Gaussian statistics, starting with a low-order non-Gaussian set of statistics, the quantile function QF, using the methodology introduced in Sect.~\ref{SubSec_methodo_comparing}. As shown in Fig.~\ref{fig:gaussStat_vs_pdf}, this first confrontation on observational data already underlines a strong degeneracy level for both Gaussian and QF features, showing that neither statistics is by itself sufficient, according to the compatibility diagnostic we introduced. In particular, the quantile statistics extract a significant amount of information that cannot be efficiently captured by Gaussian statistics.

\begin{figure}[htbp]
{\includegraphics[width=\linewidth]{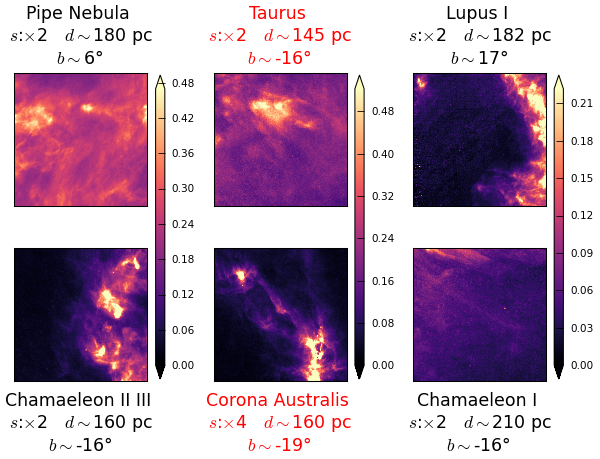}}

\caption{Examples of Gaussian confusions. \EA{Three} pairs of $512\times 512$ patches are chosen, whose locations on the scatter plot of Fig.~\ref{fig:gaussStat_vs_pdf} are given there by the red stars. The column density maps are shown in units of $10^{20}\text{cm}^{-2}$. For each patch, we report: $s$ the subsampling factor from the original 3\arcsec/pix map, $d$ and $b$ the approximated distance and Galactic latitude of the cloud. The pixel size (in $\mpc$) of a patch is thus proportional to $s \times d$. If a pair has patches $(i,j)$ with incompatible pixel sizes, that we define according to the following criterion $\max \{\frac{s_id_i}{s_jd_j}, \frac{s_jd_j}{s_id_i} \} \geq 3/2$, we color its labels in red.}
\label{fig:gaussStat_vs_pdf-1}
\end{figure}

\begin{figure}[htbp]
{\includegraphics[width=\linewidth]{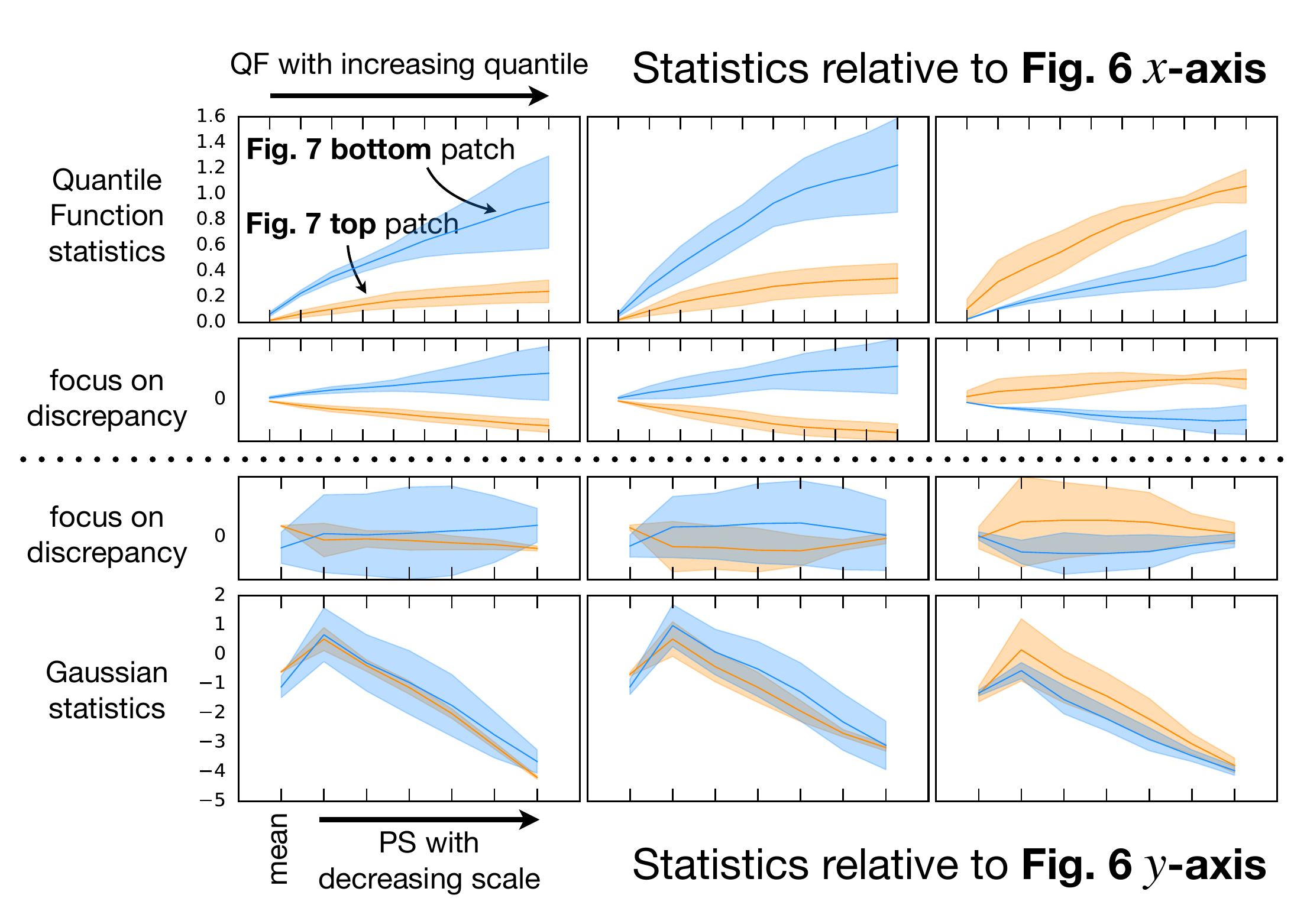}}

\caption{Statistics for the examples of Gaussian confusions shown in Fig.~\ref{fig:gaussStat_vs_pdf-1}. In each row, the orange filled line (resp. band) corresponds to the mean (resp. standard deviation) of the statistics computed over the four $256\times 256$ subpatches of the top patch of each pair of Fig.~\ref{fig:gaussStat_vs_pdf-1}, and the corresponding blue lines and areas refer to the bottom patch of the pair. The top row corresponds to the statistics used in the $x$-axis of the scatter plot of Fig.~\ref{fig:gaussStat_vs_pdf}, i.e., QF statistics, plotted with ten increasing quantile values, while the bottom row corresponds to the $y$-axis, i.e., Gaussian statistics, plotted starting with mean and followed by the binned PS with six decreasing scales. To better highlight the discrepancies between the two patches of a given pair, we report in second and third rows the offsets of the orange and blue filled lines with respect to their common mean.}
\label{fig:gaussStat_vs_pdf-2}
\end{figure}

In addition to this dataset-wide diagnostic, we display in Fig.~\ref{fig:gaussStat_vs_pdf-1} \EA{three} randomly selected pairs of $512\times 512$ patches degenerate for Gaussian statistics but with increasing QF incompatibility. The locations of these pairs on the scatter plot of Fig.~\ref{fig:gaussStat_vs_pdf} are given there by the red stars. For these \EA{three} pairs of patches, we show in Fig.~\ref{fig:gaussStat_vs_pdf-2} the two sets of statistics used in the diagnostic: QF statistics and Gaussian statistics.

\begin{figure*}[h!]
    \centering
    \subfigure[]{\includegraphics[width=.29\linewidth]{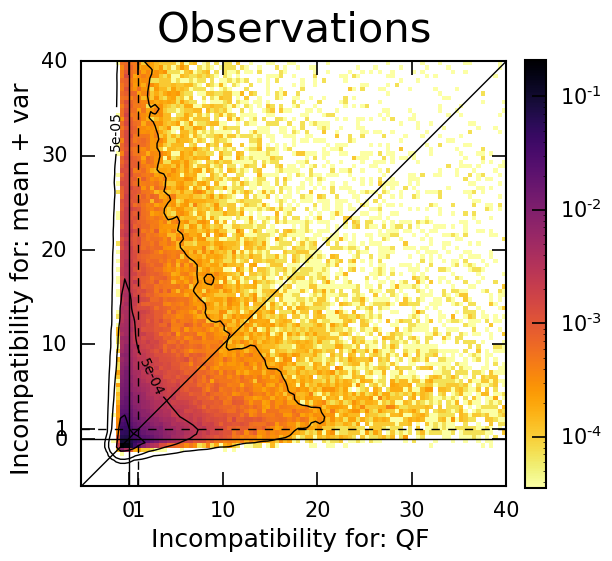}}
    \subfigure[]{\includegraphics[width=.29\linewidth]{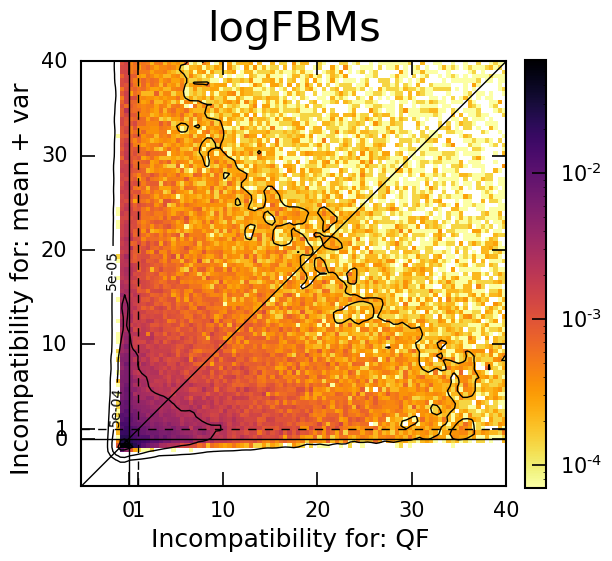}}
    \subfigure[]{\includegraphics[width=.29\linewidth]{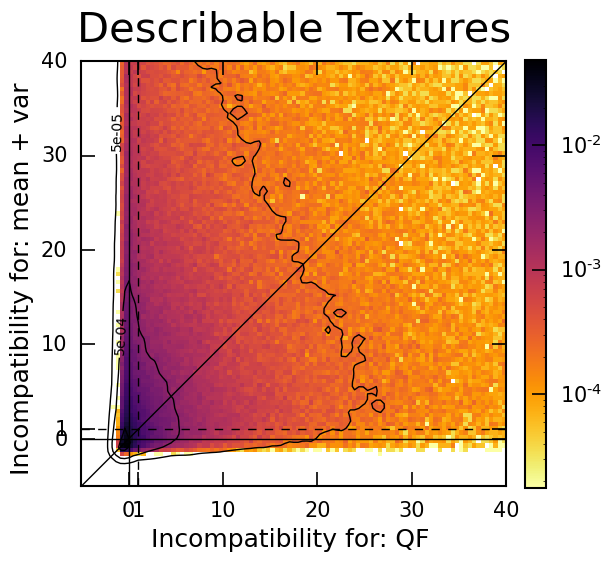}}
    \subfigure[]{\includegraphics[width=.29\linewidth]{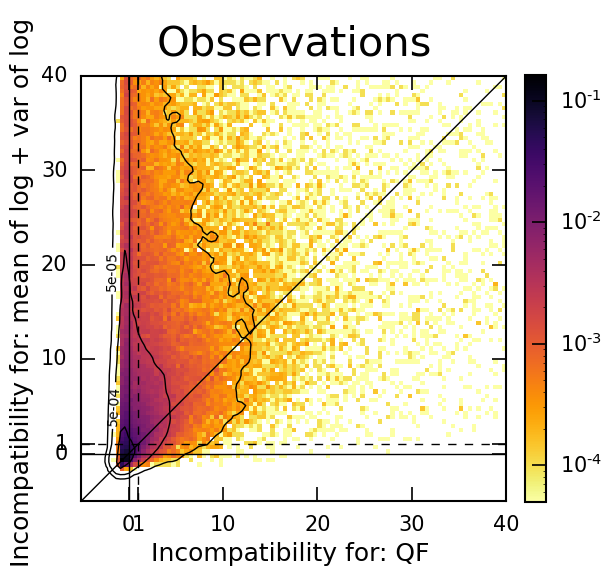}}
    \subfigure[]{\includegraphics[width=.29\linewidth]{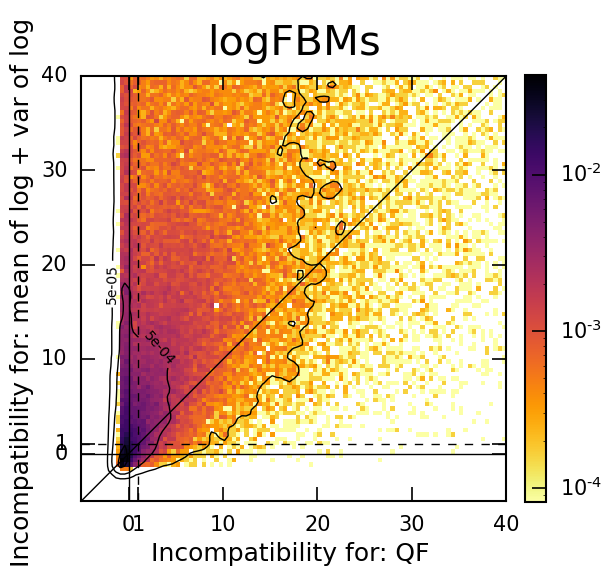}}
    \subfigure[]{\includegraphics[width=.29\linewidth]{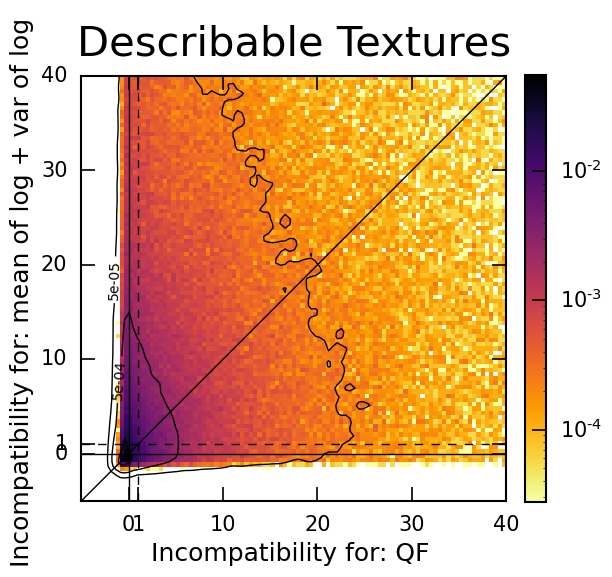}}
    
    \caption{Top row: Plots confronting QF statistics with $\{\phi_{\text{mean}},\,\phi_{\text{var}}\}$. Bottom row: Plots confronting QF statistics with $\{\phi_{\text{mean of log}},\,\phi_{\text{var of log}}\}$. These results evidence that taking the logarithm of the map enhances the discriminative power of the mean and variance statistics on both observation and logFBM data (left and middle columns) but not on DTD (right column).}
    \label{fig:loggaussStat1_vs_pdf}
\end{figure*} 

In Fig.~\ref{fig:gaussStat_vs_pdf-2}, the orange filled lines (resp. bands) correspond to the mean (resp. standard deviation) of the statistics computed over the four $256\times 256$ subpatches of the top patch of each pair of Fig.~\ref{fig:gaussStat_vs_pdf-1}, and the corresponding blue lines and areas refer to the bottom patch of the pair. To better highlight the discrepancies between the two patches of a given pair, the offsets of the orange and blue filled lines with respect to their common mean are also shown in the middle rows of Fig.~\ref{fig:gaussStat_vs_pdf-2}.

These results illustrate the ability of QF statistics to distinguish between images with the same Gaussian statistics, which confirms that our compatibility diagnostic works as expected.
We emphasize here that the $\phi_{\text{QF}}$ statistics are complementary to the one-point properties probed by Gaussian statistics. 

This behavior is not particularly surprising as one-point properties of column density maps of MCs are expected to be at least log-normal, if not with a power-law tail. Hence, estimating such properties directly from the logarithm of the maps, as does $\phi_{\text{QF}}$ but not $\phi_{\text{Gaussian}}$, might enhance the discriminative power. To test this idea, we study in the first column of Fig.~\ref{fig:loggaussStat1_vs_pdf} the degeneracy level of the set of mean and variance statistics estimated respectively on the raw maps and their logarithm. More precisely, top row plots confront QF statistics with $\{\phi_{\text{mean}},\,\phi_{\text{var}}\}$, while bottom row plots confront QF statistics with $\{\phi_{\text{mean of log}},\,\phi_{\text{var of log}}\}$. 
We see in this figure that the pairs of observations are clearly better discriminated when using the mean and variance computed on the logarithm of the maps (d) rather than directly (a). This effect is very well reproduced by logFBM data (middle column), but does not hold in general, as shown with the DTD (right column). 

These results suggest that the specific Gaussian degeneracies evidenced in Fig.~\ref{fig:gaussStat_vs_pdf} are mainly explained by the inefficiency of Gaussian one-point statistics $\{\phi_{\text{mean}},\,\phi_{\text{var}}\}$ to characterize one-point properties of processes that have a log-normal (or heavier tail) nature. 

Surprisingly, this analysis also shows that, for datasets such as observations or logFBMs, probing the PDF properties through the prism of a mean and a variance is more discriminative than probing its shape. Indeed, in both plots (d) and (e) of \EA{Fig.~\ref{fig:loggaussStat1_vs_pdf}}, our suitably constructed set $\{\phi_{\text{mean of log}},\,\phi_{\text{var of log}}\}$, of dimension two, performs almost always a better discrimination than the set $\phi_{\text{QF}}$, of dimension ten. This emphasizes the importance of suitably constructed low-dimensional descriptions in such analysis.
The effectiveness of such low-dimensional features \EA{in the diagnostic we introduced} reflects their ability to exploit underlying regularities in the processes at play, enabling data compression with only moderate loss of information.
However, when dealing with datasets made of a wide diversity of irregular processes, such as everyday life textures (e.g., DTD), it is hard to identify a compression that does not lead, for some pairs, to poorer performances (c, f).

\subsection{Molecular clouds have log-Gaussian degeneracies}

\begin{figure*}
    \centering
    \subfigure[]{\includegraphics[width=.24\linewidth]{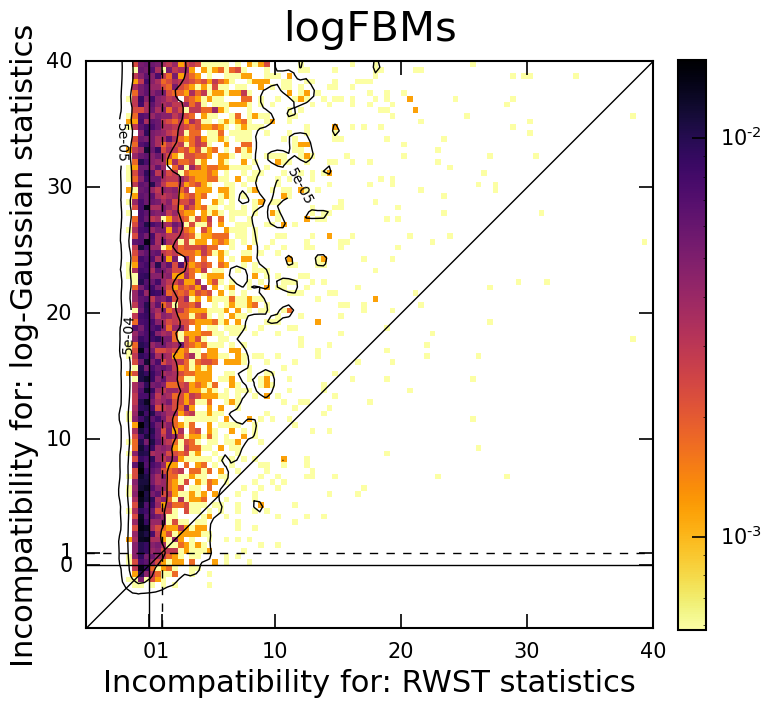}} \subfigure[]{\includegraphics[width=.24\linewidth]{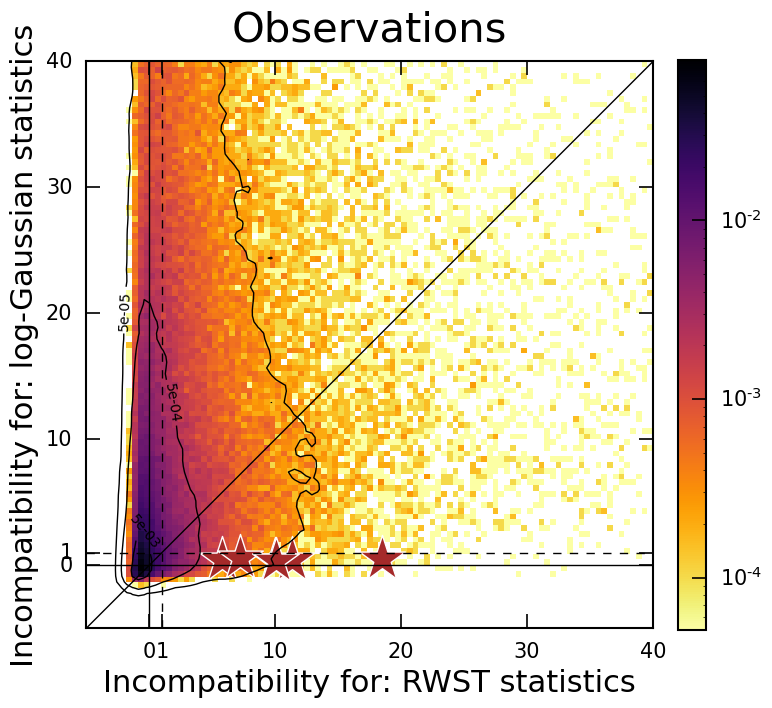}}
    \subfigure[]{\includegraphics[width=.24\linewidth]{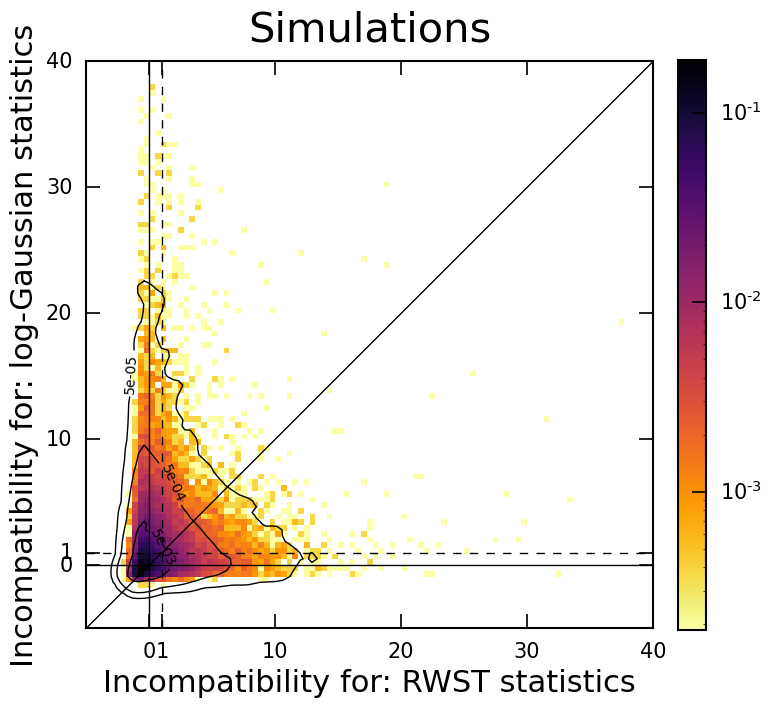}}
    \subfigure[]{\includegraphics[width=.24\linewidth]{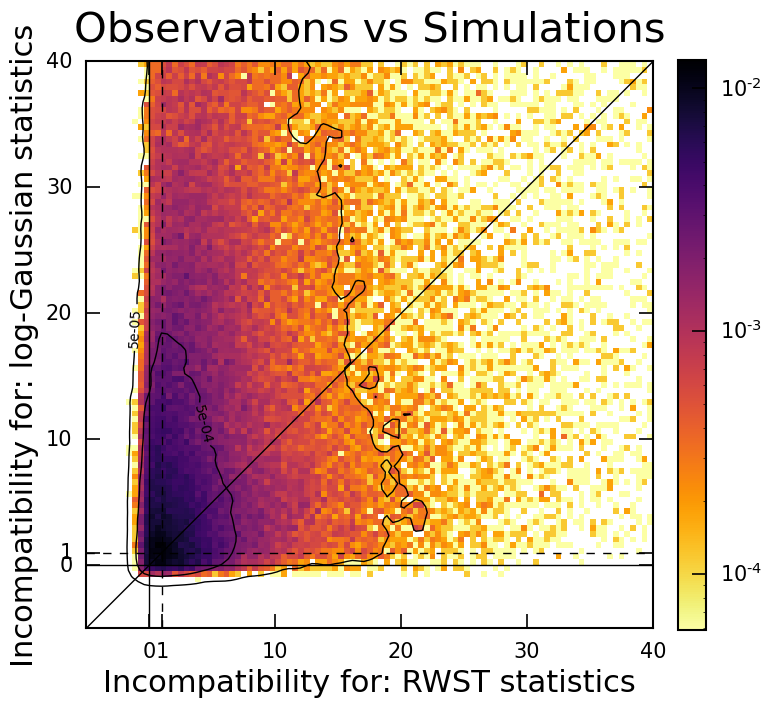}}
    
    \caption{Confrontation of log-Gaussian statistics with RWST statistics on logFBM synthetic data (a), observations (b), simulations (c), and observation-simulation pairs (d). As expected, no log-Gaussian degeneracies are found for the logFBM data (a). However, some are found for the other cases (b,c,d). To investigate the log-Gaussian degeneracies in the observations, six pairs of patches, whose locations on the scatter plot (b) are given by the red stars, are shown in Fig.~\ref{fig:loggaussStat_vs_RWST_OBS-1}. Only five stars are actually visible, but two of them are overlapped at the horizontal coordinate $x=10$.}
    \label{fig:loggaussStat_vs_RWST}
\end{figure*} 

The ability of Gaussian one-point statistics to grasp efficiently one-point properties of observations from the logarithm of the column density maps, shown in Fig.~\ref{fig:loggaussStat1_vs_pdf}, suggests to shift toward log-Gaussian statistics, that is, mean and PS estimated on the logarithms of the maps. We thus investigate now whether we may point out some degeneracies of these statistics on observational data. However, if we confront this set with QF statistics, as we did previously for Gaussian statistics, we do not expect to lift significant log-Gaussian degeneracies. Instead, we suggest searching for such degeneracies using a \EA{set of higher-order statistics: the RWST. We confront, in Fig.~\ref{fig:loggaussStat_vs_RWST}, log-Gaussian statistics with the RWST statistics, on three datasets: logFBM in (a), observations in (b) and simulations in (c).

For logFBM data (Fig.~\ref{fig:loggaussStat_vs_RWST}.a), we see that the RWST diagnostic does not provide any additional information beyond what is already captured by log-Gaussian statistics. Indeed, as expected for such a dataset, log-Gaussian statistics are sufficient~\citep{thomas2006elements}. This result, without being a complete validation, is a clear success of our compatibility diagnostic and supports its relevance.
We note that the authors do not include one-point statistics in the set of RWST coefficients, which explains why this descriptor is extremely degenerate for this dataset.

When applying the same analysis on observation data (Fig.~\ref{fig:loggaussStat_vs_RWST}.b), we see again that the log-Gaussian statistics lift strong RWST degeneracies. However, contrary to logFBM data, we now see that the RWST also lift some degeneracies of the log-Gaussian statistics.} 
To investigate more these degeneracies, we show\EA{, in Appendix~\ref{appendix:further_degeneracy_examples},} six examples of degenerate pairs in Fig.~\ref{fig:loggaussStat_vs_RWST_OBS-1} and the statistics of these patches in Fig.~\ref{fig:loggaussStat_vs_RWST_OBS-2}. 
In most of these pairs, the $S_1^{Iso}$ coefficients heavily contribute to lift the log-Gaussian degeneracy. In the first and last pairs, $\langle S_2^{Iso2} \rangle_{j_2-j_1}[\delta]$ and respectively $\langle S_2^{Iso1} \rangle_{j_2-j_1}[\delta]$ are also very discriminative.

Finally, simulations yield yet another result (Fig.~\ref{fig:loggaussStat_vs_RWST}.\EA{c}). Although this case leads to the same qualitative conclusions as observations, namely underlining the insufficiency of log-Gaussian statistics, it differs from it quantitatively. 
\EA{This shows that the distribution of simulations in the space of summary statistics is different from that of observations. In particular, the scatter plots (b) and (c) of Fig.~\ref{fig:loggaussStat_vs_RWST} show that the structures of these distributions, that is, the dependency between the statistics for each of them, are not the same.} This supports the caveat of simulation-based inference according to which observations and simulations of such \EA{ISM} processes \EA{have different statistical structures}, and motivates the observation-based approach of this paper.


\EA{To complement the previous results, we show the scatter plot obtained by directly comparing patches from observations and simulations (Fig.~\ref{fig:loggaussStat_vs_RWST}.d). This plot reveals even greater degeneracies than those observed previously, further highlighting the difference between observations and simulations. We note that this also leaves open the question of whether the set of log-Gaussian + RWST statistics also has degeneracies of its own to make such a comparison. This could be further investigated by confronting it with other sets of statistics using the diagnostic introduced in this paper.}

\section{Comparing pairs and datasets}\label{sec_distance}

In the previous section, we confronted multiple sets of statistics to assess their information content through their level of degeneracy. In this section, we fixed the set of statistics: $\phi=\phi_{\text{final}}$, composed of 17 coefficients (seven log-Gaussian descriptors and ten RWST statistics), and used it to define a \EA{morphological} distance between maps. We illustrate this distance by \EA{exhibiting} closest pairs of images in a dataset, as well as between different datasets such as observations and simulations.

\begin{figure*}
    \centering
    \includegraphics[width=\linewidth, clip]{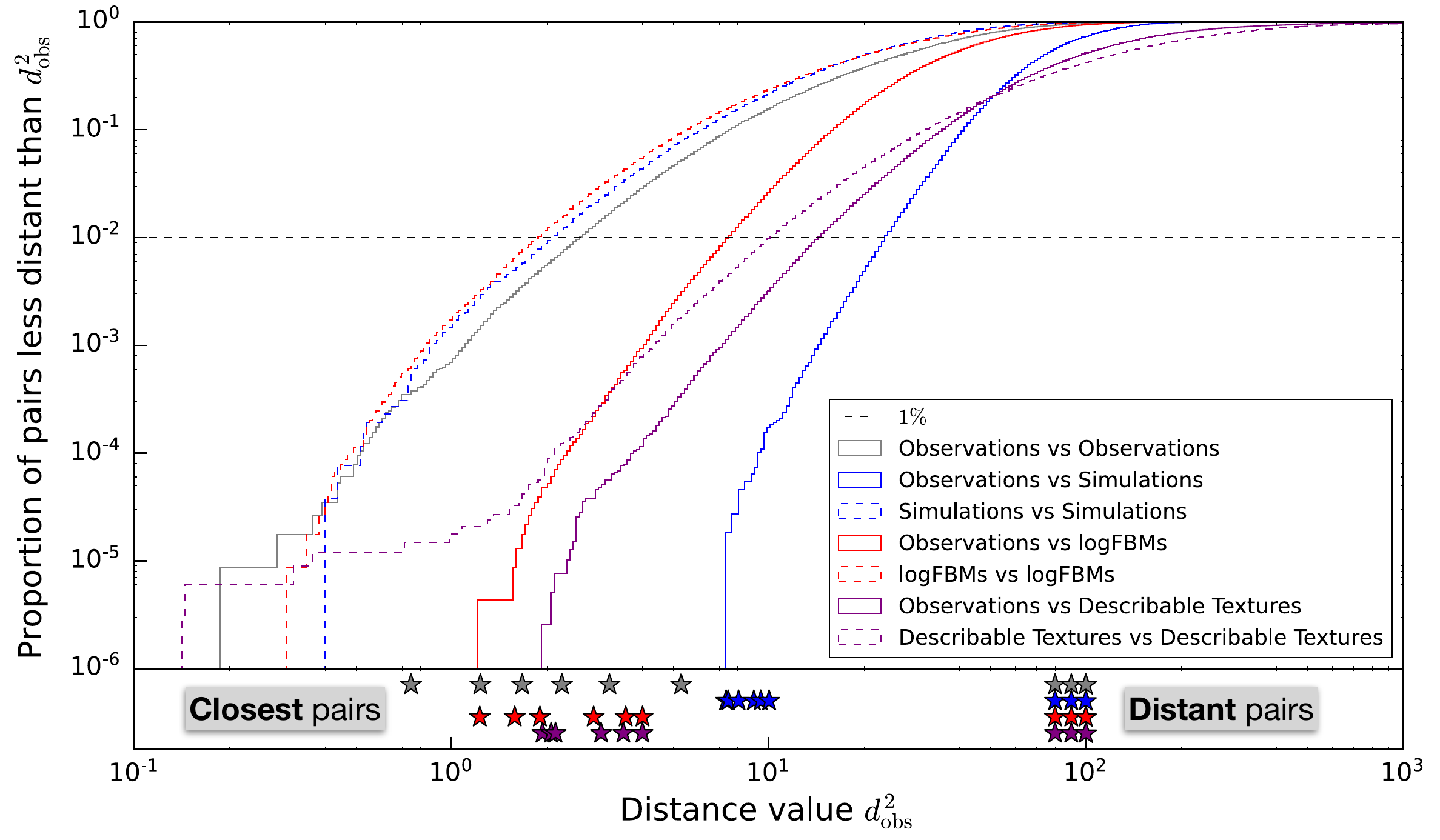}
    \caption{Cumulative distributions of $d^2_{\text{obs}}$ \EA{morphological} distances between pairs extracted from multiple datasets. The different curves correspond to different choices of datasets from which the two patches of a given pair are extracted. This distance is based on $\phi_{\text{final}}$, a set of 17 coefficients (seven log-Gaussian descriptors and ten RWST statistics). The same metric $M_\text{obs}$ is used for all distances and is defined based on the observational dataset. The stars in the bottom sub-panel correspond to "close" and "distant" pairs shown in Fig.~\ref{fig:best_match_globalVar_loggaussStat_and_RWST}. An horizontal line at the value $1\%$ is drawn to represent the proportion of pairs \EA{of neighboring patches in the sky. These pairs made of patches with strong statistical dependence are likely to lead to an underestimated distance with respect to the other pairs. While these corrupted pairs might significantly affect the gray curve below this horizontal line, they can only have a negligible influence above that line.} Same applies for pairs of simulations (blue dashed curve).}
    \label{fig:dist_distribs_globalVar}
\end{figure*}

\subsection{Defining a morphological distance}

Our objective is to define a \EA{morphological} distance between two patches based on $\phi_{\text{final}}$. One of our requirements is to enable a comparison between distance values for different pairs, so that these pairs can be ordered according to the morphological proximity of their patches. This requirement prevents us from using the statistical compatibility diagnostic introduced earlier. Indeed, since it is weighted by the local variance of each patch's statistics, this can lead, for example, to some patches in a dataset being compatible with almost all the others simply because their spatial variance is very high. This could also encourage a simulation to approach an observation, following this criterion, by arbitrarily increasing its variance $\operatorname{Var} \phi(x_\text{SIM})$ without focusing on reducing the discrepancy of its average statistical properties $\hat \phi(x_\text{SIM})-\hat \phi(x_\text{OBS})$. This property was purposely used to act as a penalization when confronting different sets of statistics in the previous section, but is no longer desired for pairs' ordering, once $\phi$ is fixed.

To build a \EA{morphological} distance that avoids this drawback, we choose instead to normalize it by the variability of the statistics evaluated over the entire observational dataset. We thus use the following distance:
\begin{equation}
d^2_{\mathcal{D}}(x_i,x_j) \equiv (\hat \mu_i-\hat \mu_j)^T(\operatorname{diag} M_{\mathcal{D}})^{-1} (\hat \mu_i-\hat \mu_j),
\end{equation}
which is normalized by the \EA{variance} of $\hat \mu_i$ over all maps of a given dataset $\mathcal{D}$:
\begin{equation}\label{eq_metric_def}
M_\mathcal{D} \equiv \big\langle (\hat \mu_i - \langle \hat \mu_j\rangle_j)(\hat \mu_i - \langle \hat \mu_j\rangle_j)^T \big\rangle_i,
\end{equation}
where the brackets indicate an average over $\mathcal{D}$. We note, however, that it is difficult to interpret the value of $d^2_{\mathcal{D}}$ in absolute terms. Indeed, the $M_\mathcal{D}$ term does not describe a typical variance for a given process, but describes the variety of morphologies encountered accross the entire dataset, that can be broad, as seen in the MC data investigated here. Unlike statistical compatibility diagnostics, the $d^2_{\mathcal{D}}$ distance is therefore modified by the addition or removal of maps in the dataset $\mathcal{D}$, and can be affected by the presence of outliers.

In the following, we work with different datasets. For instance, we aim at comparing the minimum distance between observations and simulations to the typical distance of the closest pairs of observations. To do so, we use in this paper a unique metric, $M_\text{obs}$, computed on the observation dataset, as well as its associated $d^2_\text{obs}$ distance.

\subsection{Closest pairs}

\begin{figure*}
    \centering
    \includegraphics[width=\linewidth]{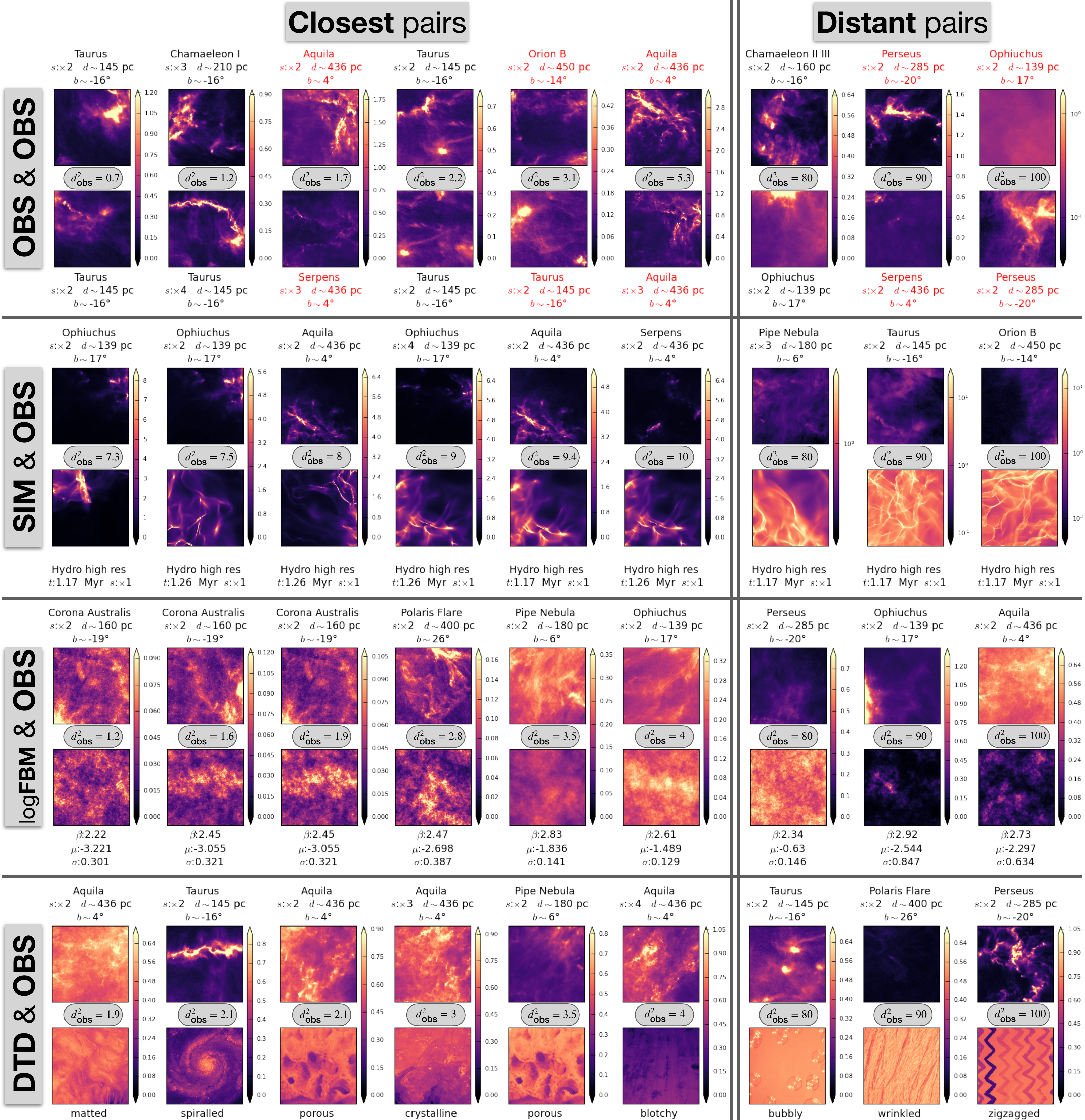}
    \caption{Closest (left) and more distant (right) pairs extracted from distributions of \EA{morphological} distances reported in Fig.~\ref{fig:dist_distribs_globalVar}. Column density maps are shown in units of $10^{20}\text{cm}^{-2}$. The colorbars can change from a pair to another. We see that the closest (OBS, SIM) pairs are much more distant ($d^2_\text{obs}\sim 7$) than the closest (OBS, OBS) pairs ($d^2_\text{obs}\sim \EA{0}.7$). See Sect.~\ref{SubSecIntObsSim} for more detailed interpretation on these results. Some logFBM models end up quite close to the most diffuse regions observed: Polaris Flare and Corona Australis. Such regions are highly contaminated by CIB emission. Many (OBS, DTD) pairs are found to be close whereas the textures look very different. This shows that the set of summary statistics developed in this paper is tuned for ISM observations but is far from being sufficient for any kind of data. This also illustrates that MCs have much more regularity in terms of morphology than DTD textures.}
    \label{fig:best_match_globalVar_loggaussStat_and_RWST}
\end{figure*}

We use the $d^2_\text{obs}$ \EA{morphological} distance to identify the closest pairs of patches that can be found between two datasets. To do so, we report in Fig.~\ref{fig:dist_distribs_globalVar} the cumulative distributions of distances associated with pairs of observation patches (in gray), as well as with pairs consisting of one observation patch and one logFBM, DTD, and simulation patch (in red, purple, and blue, respectively). For each case, we select six of the typical closest pairs\footnote{\label{footnote_obs_pairs_bias}We do not investigate the $\sim 1\%$ closest pairs of observations, since the comparison between observations is slightly biased with respect to the other ones. The reason is that, in that case only, we have pairs made of non-independent patches, such as neighboring patches in the sky or two slightly different scaled versions of the same region. This corresponds to approximately $1\%$ of the pairs, which we therefore exclude.}, as well as six relatively distant pairs, that we show in~Fig.~\ref{fig:best_match_globalVar_loggaussStat_and_RWST}. In dashed lines with the same colors, we also report the cumulative distributions of distances associated to pairs of patches from a common dataset (logFBM, DTD\footnote{We note that the $\sim10$ closest pairs of DTD images, located at $d^2_\text{obs}\lessapprox1$ are artifacts in this dataset corresponding in practice to almost identical images.}, and simulation patches, respectively).

Closest (OBS, OBS) pairs and (OBS, logFBM) pairs are visually rather similar, while distant pairs look very different. This relative agreement between the morphological proximity, as probed by our statistical distance and by human vision, is encouraging, because it should be satisfied by an ideal distance. However, it remains far from being an exhaustive diagnostic.

The closest (OBS, logFBM) pairs have a distance $d^2_\text{obs}\sim 2$, which corresponds to a high agreement. Indeed, only $\sim 0.5\%$ of (OBS, OBS) or (logFBM, logFBM) pairs exhibit smaller distance values, and we have already mentioned$^{\ref{footnote_obs_pairs_bias}}$ that the closest 1\% of (OBS, OBS) pairs are very similar by construction. In comparison, simulations are farther from observations: the closest pairs between observations and simulations have a distance $d^2_\text{obs}\sim 8$, four times larger than the closest (OBS, logFBM) pairs, when already $\sim 10\%$ of pairs of observations exhibit smaller distances (see Sect.~\ref{SubSecIntObsSim} for more details on this interpretation). These closest (OBS, SIM) pairs \EA{seem to us} visually less similar than the closest (OBS, OBS) or (OBS, logFBM) pairs (left panels of Fig.~\ref{fig:best_match_globalVar_loggaussStat_and_RWST}), but are still not so different, for instance with respect to the more distant pairs, that can be seen in the right panels of Fig.~\ref{fig:best_match_globalVar_loggaussStat_and_RWST}. 

\begin{figure*}
    \centering
    \includegraphics[width=\linewidth]{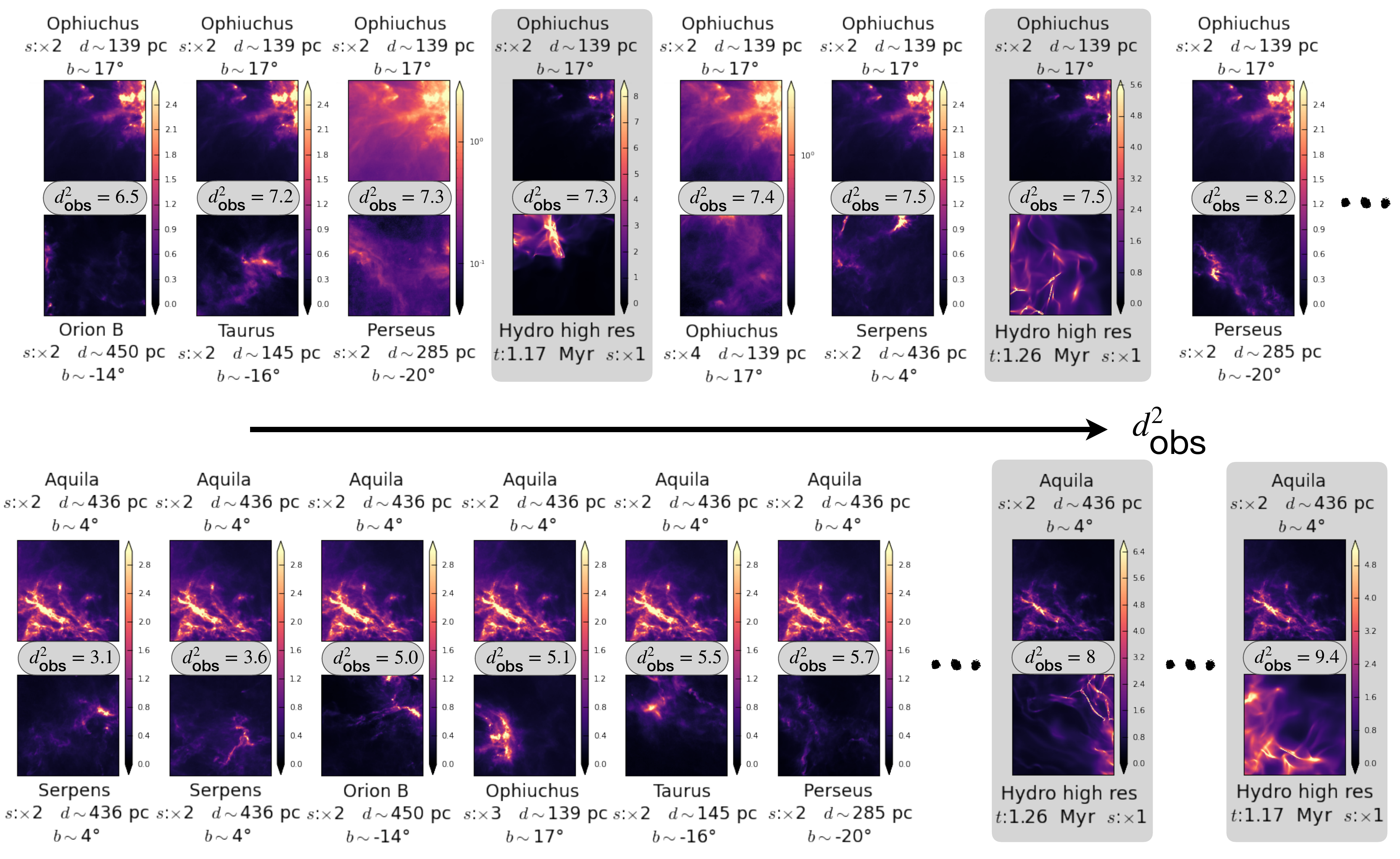}
    \caption{Comparison of closest (OBS, SIM) pairs $(i,j)$ (gray boxes) to neighboring (OBS, OBS) pairs $(i,i')$. The top row focuses on the closest observation to simulations, which corresponds to a patch of Ophiuchus, while the bottom row focuses on the second closest patch, which is in Aquila. In the latter case, neighboring (but independent) observations of Aquila are significantly closer than the closest simulations. The maps are shown in units of $10^{20}\text{cm}^{-2}$ but the colorbars can change from a pair to another.}
    \label{fig:comparing_close_obs_sim}
\end{figure*}

The closest (OBS, logFBM) samples obtained here are very diffuse regions of MCs such as Corona Australis, Polaris Flare, and Ophiuchus. 
This is not surprising because, for MCs, diffuse regions are closer to logFBM models than dense regions, whose PDFs are known to deviate from log-normality.  
However, these diffuse regions are still supposed to exhibit coherent structures that should induce deviations from logFBM models. Here, such deviations are found to be small. We note, however, that this proximity only means that certain diffuse regions are close to logFBM processes relative to the total variability of the observational dataset, and not necessarily that they are well described by such models, or could not be distinguished by the previous compatibility diagnostic. Moreover, this proximity to logFBM models is also partly due to contamination by the \EA{CIB} which has non-negligible power at such low levels of column density, and is expected to Gaussianize the data, including their RWST statistics~\citep{auclair2024separation}. Incidentally, we remind that the observational dataset has some artifacts that make it deviate from an "ideal" MC column density dataset, even if we try to limit as much as possible their effects (noise, finite resolution), as discussed in Sect.~\ref{sec_statistics}. 

On the contrary, the observations that are closest to the simulations of dense MCs correspond, unsurprisingly, to dense regions of MCs such as Ophiuchus, Aquila, and Serpens. We note that Ophiuchus exhibits patches that are close to these dense simulations, but also at least one patch that is close to a logFBM model, underlining the spatial variability of molecular cloud morphologies. This illustrates the difficulty to treat a MC as a single entity, and emphasizes the relevance of our local approach. 

The comparison between observations and DTD shows the limitations of our \EA{morphological} distance diagnostic. Indeed, the closest (OBS, DTD) pairs are found at a distance $d^2_\text{obs}\sim 2$ that is the typical distance between close (OBS, OBS) pairs or close (OBS, logFBM) pairs, although they are visually very different. This illustrates that, for a diagnostic based on a low-dimensional set of statistics, it is difficult to probe a distance over a family of processes that has such a wide variety of textures as DTD. On the contrary, because the simulations closest to the observations are more distant ($d^2_\text{obs}\sim 8$), this suggests that the set of simulations does not intersect the set of observations and that $\phi_\text{final}$ is able to pinpoint this discrepancy, as discussed in the following subsection.

Finally, it is quite impressive to see that we can build a distance diagnostic from a representation of dimension 17 only that still manages to identify morphological similarity between maps quite satisfactorily. This illustrates the possibility of constructing a highly informative but low-dimensional description tailored to a family of processes from an ensemble of compressed sets of typical statistics. It should be stressed, however, that this study remains partial, notably because it is based mainly on the observation of a few close pairs in our dataset. Yet, the confrontation diagnostics studied in Sect.~\ref{section results} showed that degenerate counterexamples remain largely in the minority. In addition, we lack solid baselines since it is inherently difficult to quantify visual impressions of morphological proximity, although some work has been done in this direction~\citep{Peek2021Search}.

\subsection{Interpreting the minimal distance between observations and simulations}
\label{SubSecIntObsSim}

A last question we tackle is whether the relatively high value of the minimal distance $d^2_\text{obs}$ that we get between observations and simulations indeed indicates \EA{that these two sets of processes do not overlap. To do so, we select the observations that are closest to the simulations. As shown in Fig.~\ref{fig:best_match_globalVar_loggaussStat_and_RWST}, these correspond to patches of Ophiuchus and Aquila. Then, we report in Fig.~\ref{fig:comparing_close_obs_sim} the closest observations to these patches, and sort them by proximity, according to $d^2_\text{obs}$.
\begin{itemize}
    \item Concerning the Ophiuchus patch (top row), only three observations patches seem to be closer ($d^2_\text{obs} \in \{6.5,7.2,7.3\}$) to it than the closest simulation ($d^2_\text{obs} =7.3$), and the difference in $d^2_\text{obs}$ is small. 
    \item However, concerning the Aquila patch (bottom row), there are numerous observations closer to it than the closest simulation ($d^2_\text{obs} =8$), and their distance is much smaller (down to $d^2_\text{obs} =3.1$).
\end{itemize}
}

We further interpret this discrepancy between distances in Appendix~\ref{appendix:mitigating_bias_distance}. We conclude that the minimum (OBS, SIM) distance value, obtained on all possible pairs between those datasets, evidences a meaningful but moderate distinction between these datasets. We note that the $\phi_\text{final}$ statistics set may still have some degeneracies, particularly for the (OBS, SIM) comparison, and that accounting for these should likely deepen this gap. However, we believe that the current analysis illustrates the usefulness of such a \EA{morphological} distance, and leave a more detailed study for future work.


\section{Conclusions} \label{conclusion}

In this paper, we aim to study the diversity of morphologies of observed molecular clouds. To do so, we construct a set of $\sim 500$ patches of size $512\times 512$ pixels, extracted at different resolutions from column density maps of 14 nearby clouds derived from the HGBS dust emission observations. We compute several sets of statistics (mean, variance, quantile function, power spectrum and scattering transform) from these maps, and we compare their informative power. To do so, we introduce a new methodology (Fig.~\ref{fig:illustration_test}) that allows us to confront, without any supervision, two sets of summary statistics on their respective abilities to detect statistical incompatibility between pairs of \EA{stochastic} processes \EA{of} a given \EA{family}. 

Applying this methodology to this set of observations, we find that Gaussian statistics have degeneracies for the observational dataset, some of which can be lifted by one-point statistics (Fig.~\ref{fig:gaussStat_vs_pdf}). We then show that even if log-Gaussian statistics are much less degenerate, a compressed set of scattering statistics still succeeds to demonstrate further degeneracies (Fig.~\ref{fig:loggaussStat_vs_RWST}\EA{.b}). This confirms that the diversity of morphologies arising in these observed clouds cannot be sufficiently described by either Gaussian or log-Gaussian statistics. This means that using such descriptions, typically to compare numerical simulations with observational data, can lead to misleading \EA{conclusions}: in addition to missing potential absolute discrepancies, observations with different \EA{statistical} properties can still be matched to the same simulation.

We apply the same diagnostic to simulations \EA{(Fig.~\ref{fig:loggaussStat_vs_RWST}.c)} and to the set of logFBMs (Fig.~\ref{fig:loggaussStat_vs_RWST}\EA{.a}), and find deviations from the \EA{statistical structure} of the set of observations. In particular, the outcome of this diagnostic strikingly supports the sufficiency of log-Gaussian statistics to discriminate between logFBM patches. Regarding observations and simulations, there remain deviations between their respective \EA{statistical structures}, even though great care has been taken in this work to design robust and low-dimensional sets of summary statistics. This supports the difficulty of transferring statistical properties \EA{learned on simulations to observations, especially those properties relative to higher-order moments, as mentioned in the caveat n°\ref{caveat_2} in the introduction}. \EA{This} also buttresses the supervision-detached approach developed in this paper, along with the choice to work with compressed and robust summary statistics. 


From these results, we introduce a morphological distance $d^2_\text{obs}$ based on a set of summary statistics $\phi_\text{final}$ composed of seven log-Gaussian and ten RWST coefficients. The similarity probed by this distance is in agreement with visual impression when comparing observations with themselves, with logFBMs and with simulations (Fig.~\ref{fig:best_match_globalVar_loggaussStat_and_RWST}). However, it remains insufficient to operate on datasets made of a wider diversity of textures such as DTD.
\\

This work opens multiple perspectives:
\begin{itemize}
    \item The methodology we developed to confront summary statistics requires very few assumptions: it can operate in an unsupervised and very low data regime. Hence, such methodology can easily be assimilated by the ISM community and applied on a wider set of statistics and physical fields (velocities, polarization, temperature).
    \item The low-dimensional and analytical morphological embedding $\phi_\text{final}$ developed here allows us for a \EA{statistically} interpretable comparison, for instance between different observed clouds, between observations and simulations or statistical models. It also paves the way for the use of more sophisticated unsupervised learning techniques.
    \item Leveraging saliency maps $\nabla_\text{pixels} d_\text{obs}^2(x_i, x_j)$, the \EA{morphological} distance $d^2_\text{obs}$ can be used to highlight the main areas responsible for morphological discrepancies between two patches $(x_i,x_j)$, broadening the scope of the work initiated by \cite{peek2019androids} to the unsupervised world of observations. 
    \item The present work focuses on the statistical properties of molecular clouds. In particular, it assumes that patches with incompatible statistics should be considered as realizations of different processes. However, a potential caveat of this approach, when applied to nonstationary processes such as dense molecular clouds, could be to find statistical incompatibilities between patches which are not particularly relevant from a physical point of view. While we do not further discuss this issue in the present paper, we think it could deserve further work.
    \item The confrontation methodology can be used to make a feature selection algorithm, in the spirit of the FRAME model developed by \cite{zhu1998filters}, but designed to optimize the comparison task of a nonparametric collection of processes $\{p_i\}_i$, instead of modeling a single process.
\end{itemize}
The following improvements could also be of great benefit:
\begin{itemize}
    \item Reduce the uncertainty in the compatibility diagnostic $d^2_\phi$ due to the precision matrix estimation. A promising idea is to use a maximum entropy model conditioned on the data\EA{~\citep{bruna2019multiscale, allys_rwst_2019, zhang2021maximum}} on which to perform the precision estimation.
    \item Reduce the overall dependency of the \EA{morphological} distance $d^2_\mathcal{D}$ on the dataset $\mathcal D$, and in particular on its outliers. This can be done by building a localized metric, \EA{reducing the scope of the averages in Eq.~\ref{eq_metric_def} from} $\mathcal{D}$ to $\mathcal{D}_\text{loc}[i]$, the \EA{set of} neighbors of a given process $i$.
\end{itemize}

\begin{acknowledgements}
\EA{We thank the anonymous referee for his careful reading of the manuscript, detailed comments and useful recommendations that helped us to improve the quality of the paper.}\\

This research has made use of data from the \textit{Herschel} Gould Belt survey (HGBS) project\footnote{\url{http://gouldbelt-herschel.cea.fr}}. The HGBS is a \textit{Herschel} Key Programme jointly carried out by SPIRE Specialist Astronomy Group 3 (SAG 3), scientists of several institutes in the PACS Consortium (CEA Saclay, INAF-IFSI Rome and INAF-Arcetri, KU Leuven, MPIA Heidelberg), and scientists of the \textit{Herschel} Science Center (HSC).\\

This work reused datasets available on the Galactica simulations database\footnote{\url{http://www.galactica-simulations.eu}}.\\

This work used the Describable Textures Dataset\footnote{\url{https://www.robots.ox.ac.uk/~vgg/data/dtd/}}.
\end{acknowledgements}

\bibliographystyle{aa}
\bibliography{references.bib}

 \begin{appendix}
 


\section{Further details on the datasets}
\label{appendix:otherdatasets}

\subsection{Observations}
\begin{figure}[h!]
    \centering
    \includegraphics[width=1\linewidth]{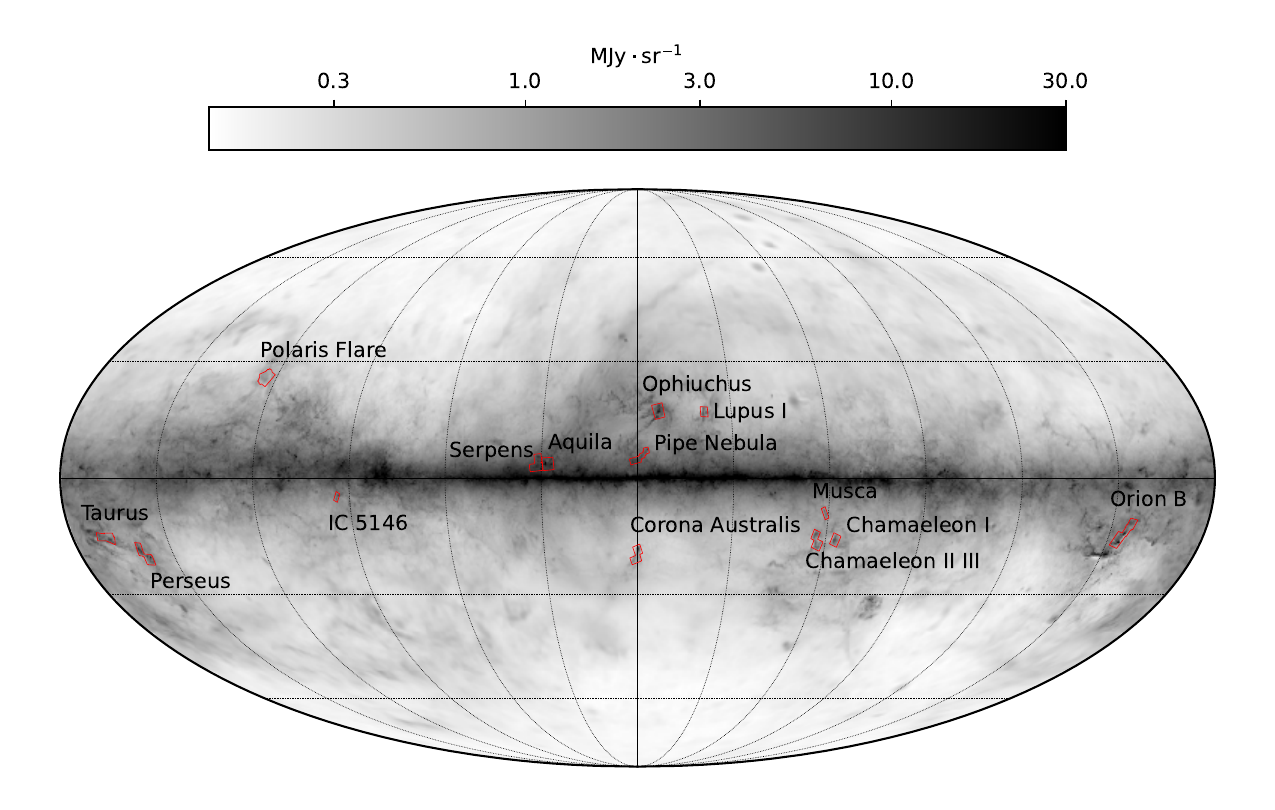}
    \caption{Footprints of the \textit{Herschel} Gould Belt Survey (HGBS) fields used in this study, overlaid on the total thermal dust intensity at $353\,\mathrm{GHz}$ from the GNILC~\citep{remazeilles2011GNILC} variable resolution data of {\it Planck}~\citep{Planck-2018-XII}.}
    \label{fig:hgbs_view}
\end{figure}

Fig.~\ref{fig:hgbs_view} presents the footprints on the sky of the set of observations used in this work.

\subsection{Numerical simulations} \label{SIM}

\begin{figure}[htbp!]
    \centering    \includegraphics[width=.8\linewidth]{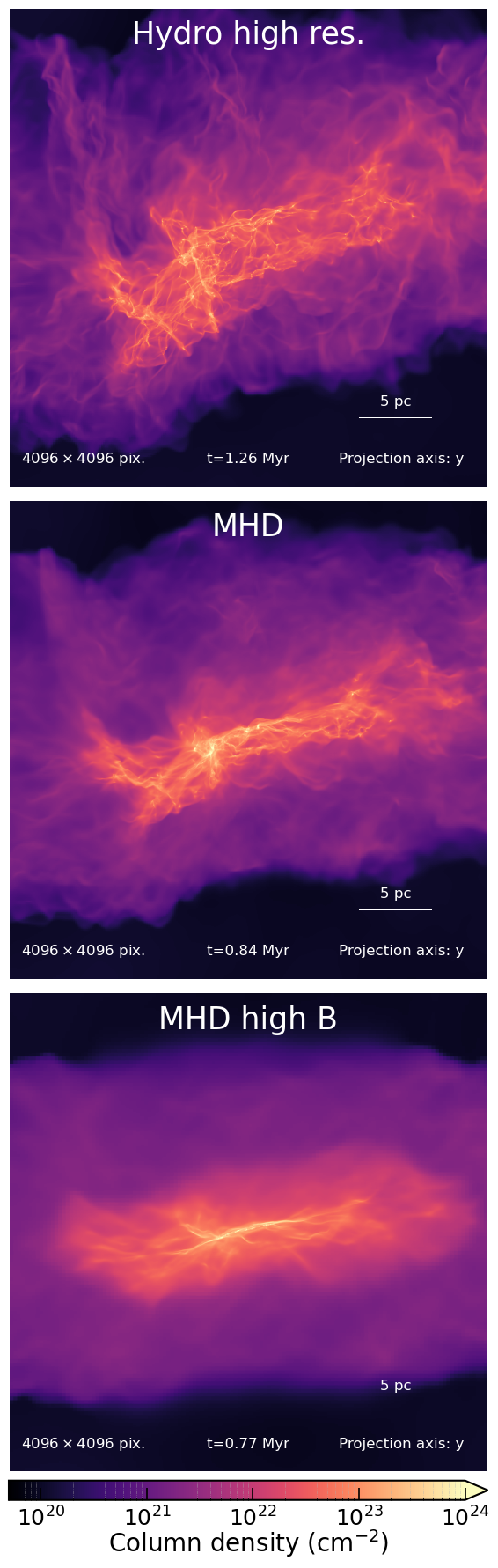}
    \caption{Overview of the simulations. One snapshot is shown for each model ("Hydro high res", "MHD", and "MHD high B"), and the maps show the central $33\,\mathrm{pc}\times 33\,\mathrm{pc}$ field for \EA{$\mathrm{H_2}$} column density integrated along the $y$ axis.
} \label{fig:sim_used}
\end{figure} 

The set of numerical simulations used in this paper as an example of state-of-the-art attempts to reproduce the physics of the ISM in silico is taken from the ORION project of the Galactica database\footnote{\url{http://www.galactica-simulations.eu/db/STAR_FORM/ORION/}}. Using the adaptive mesh refinement (AMR) code RAMSES~\citep{Teyssier2002,Fromang2006}, the data simulates the collapse of a dense molecular cloud under self-gravity, including decaying MHD turbulence but without stellar feedback. The focus of these simulations is to study the early stages of the star formation process in a molecular cloud ($10^5\,\mathrm{M}_\odot$ in a 66\,pc cubic box), and so fits in well with the HGBS observational data used in this paper.

Three classes of simulations are considered, one without magnetic field ("Hydro high res"), one with a typical magnetic field of order $5-10\,\mu\mathrm{G}$ ("MHD"), and one with a higher value of the magnetic field of order $10-25\,\mu\mathrm{G}$ ("MHD high B"). With the adaptive mesh refinement scheme, the spatial resolution of the models can reach down to 1\,mpc ($\sim 200 \au$), and even $100 \au$ for the "Hydro high res" model. For more details about these simulations, we refer the reader to~\cite{ntormousi_core_2019}.

For each of these models, we take two snapshots, integrate along the three axes, and crop the resulting column density images to keep the central $33\,\mathrm{pc}\times 33\,\mathrm{pc}$ field, on a regular $4096\times 4096$ grid. Examples of such images are shown in Fig.~\ref{fig:sim_used}. As we did for observations, we cut these panels into $512 \times 512$ patches, keeping only the ones where the effective resolution is fine enough, to avoid artifacts due to the AMR scheme affecting our morphological analysis. The pixels in the resulting patches have sizes $8\,\mathrm{mpc}$. This is in accordance with the typical spatial sampling found in our observational dataset. Indeed, this corresponds to the finest pixels (6\arcsec) of intermediate-distance clouds $d\sim 300 \pc$, but also to the closest clouds $d\sim 150 \pc$ sampled at 12\arcsec.

\subsection{logFBM models} \label{logFBM}

To perform our analysis of observational data, we also considered purely synthetic, parametric models of column density maps, derived from exponentiations of FBMs. Such fields have been previously studied by the ISM community \citep{elmegreen2002fractal, brunt2002interstellar, miville2007statistical, levrier_statistics_2018}. It is indeed a convenient class of models that allows us to simultaneously reproduce exactly the log-normal one-point properties (encountered in quiescent regions) and power-law power spectra to a good approximation. An example is given in Fig.~\ref{fig:ISM_is_not_logFBM}.

In this paper, we consider the following model:
\begin{equation}
    X \equiv e^{\sigma F_\beta \star Z + \mu}, \,\,\, \textrm{with}~Z \sim \mathcal N(\vec 0, I),
\end{equation}
parametrized by $\mu$, $\sigma$ and $\beta$, corresponding respectively to the mean, standard deviation and spectral index of the FBM. The latter is generated by sampling a $512 \times 512$ Gaussian white noise map $Z$, which is then filtered with $F_\beta$, that is defined in Fourier space as: 
\begin{equation}
    \tilde F_\beta(\vec k) \equiv 
    \begin{cases}
    \propto k^{-\beta/2} &\text{ if } \vec k \neq \vec 0,\\
    1 & \text{ if } \vec k = \vec 0,
    \end{cases}
\end{equation} 
such that $\sigma F_\beta \star Z + \mu$ is a real valued Gaussian process with mean $\mu$, variance $\sigma^2$ and $\beta$-decaying PS power-law.

\begin{figure}[htbp!]
    \centering
    \includegraphics[width=\linewidth]{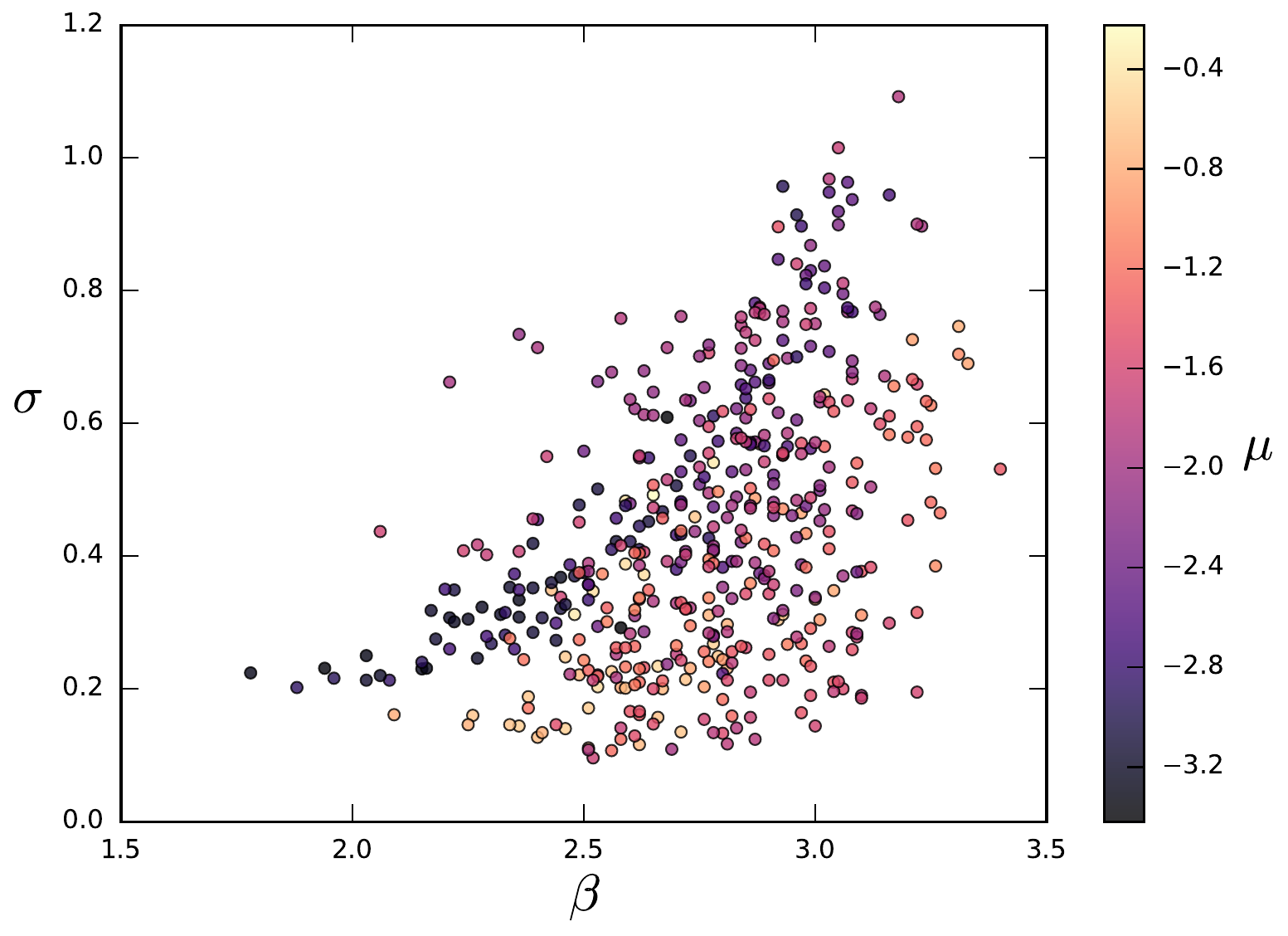}
    \caption{log-Gaussian parameters fitted on the observations used to sample logFBM data.}
    \label{fig:logFBM_params}
\end{figure}

The parameters $(\mu, \sigma, \beta)$ are fitted from the observations. More precisely, for each $512 \times 512$ column density patch that has no negative pixel (that is about $87\%$ of the 550 observational patches), we estimate the mean $\mu$, variance $\sigma^2$ and spectral index $\beta$ on the logarithm of the column density patch, expressed in $10^{22}\,\mathrm{cm^{-2}}$ units. This procedure leads to a distribution of $\sim 480$ estimated parameters triplets $(\mu, \sigma, \beta)$ that is reported in Fig.~\ref{fig:logFBM_params}. For each triplet, we then sample one $512 \times 512$ logFBM patch. Contrary to the other patches considered in this paper (observations, simulations, everyday textures), these samples have PBCs, by construction. However, their $256 \times 256$ subpatches, that are the only ones on which the various summary statistics are applied, do not, just as the rest of the data.

We dub these synthetic maps logFBM models, to recall that the statistics of their logarithms are those of FBM fields. In the literature, emission maps are often produced by integrating a 3D generated field. In this paper, for simplicity and computational efficiency, we directly generate 2D fields. Even though integrated 3D fields could be expected to better model MCs' column density maps, they do not differ much in morphology from 2D models and both are far from reproducing MCs' morphology. Thus, the 2D logFBM models used here are suited to the scope of this paper.

\subsection{Describable textures dataset} \label{DTD}
To better understand the peculiarities of MCs' statistical properties, we compare them with a very wide set of everyday textures: the DTD~\citep{cimpoi14describing}. This dataset is openly accessible\footnote{\url{https://www.robots.ox.ac.uk/~vgg/data/dtd/}}. In this paper, we keep only images with a smaller edge size larger than 512 pixels. We then crop these $\sim1050$ resulting pictures into $512 \times 512 $ patches. Next, each patch is converted to grayscale by summing its three RGB channels. These maps have integer values ranging from 0 to $3 \times 255 = 765$ that we finally divide by a factor $10^3$ to match the typical values of the other datasets.

\section{Srivastava and Du test statistic}
\label{App_TestS&D}
The test statistic $d^2_\phi(x_i,x_j)$ we introduce in Eq.~\ref{test_stat} is derived from \cite{srivastava2008test}. We apply this test on the two collections of summary statistics $\{\phi(x_i^{(l)})\}_{1\leq l \leq N_i}$ and $\{\phi(x_j^{(l)})\}_{1\leq l \leq N_j}$ computed on the subpatches (indexed by $l$) of the patch $x_i$ and the patch $x_j$. In our case, $N_i=N_j=4$. We recall Eq.~\ref{test_stat} here:
\begin{equation}
    d_\phi^2(x_i, x_j) \equiv \alpha \Big[\big( \hat\mu_i-\hat \mu_j \big)^T D_S^{-1}  \big( \hat\mu_i-\hat\mu_j \big)-\beta\Big].
\end{equation}
This equation is based on the following quantities:
$$ \hat \mu_i \equiv \langle \phi(x_i^{(l)}) \rangle_{1\leq l \leq N_i}, $$
$$ p=\text{dim } \phi, \text{ and } n=N_i+N_j-2,$$
$$\begin{cases}
    S_i \equiv \langle \big[\phi(x_i^{(l)})-\hat\mu_i \big] \big[\phi(x_i^{(l)})-\hat\mu_i \big]^T \rangle_{1\leq l\leq N_i}, \\
    S \equiv \frac{1}{n} (N_i S_i + N_j S_j), \\
    D_S \equiv \operatorname{diag} S, 
\end{cases}$$
$$\begin{cases}
    R \equiv D_S^{-1/2} S D_S^{-1/2}, \\
    c_{p,n} \equiv 1+\operatorname{tr}R^2/p^{3/2}, \\
    \alpha \equiv \frac{N_iN_j}{N_i+N_j} \frac{1}{\sqrt{2(\operatorname{tr}R^2-p^2/n)c_{p,n}}} , \\
    \beta \equiv \frac{N_i+N_j}{N_iN_j}\frac{np}{n-2}.
\end{cases}$$


\section{Rationale for using the logarithm of some standard statistics}
\label{AppLogStat}

In theory, because the mapping $\psi \mapsto \log \psi$ is invertible, the same amount of information should be contained in $\phi(x)$ and in $\log \phi(x)$ (when the latter is well defined). However, for the compatibility diagnostic we chose, these two options might perform differently. Such issues are not specific to this test and also occur with standard diagnostics such as Fisher analysis~\citep{park2023quantification}.

First, this nonlinear mapping can change the fulfillment level of the test assumptions. In the Srivastava and Du test used here, the two input distributions to be compared are assumed to be normally distributed and with equal variance. In many cases, typically when $\phi(x)>0$, taking the logarithm of $\phi(x)$ decreases the deviation to normality and decreases the relative deviations between the variances of $\phi(x_i)$ and $\phi(x_j)$.

\begin{figure}[htbp]
    \centering
    \includegraphics[width=\linewidth]{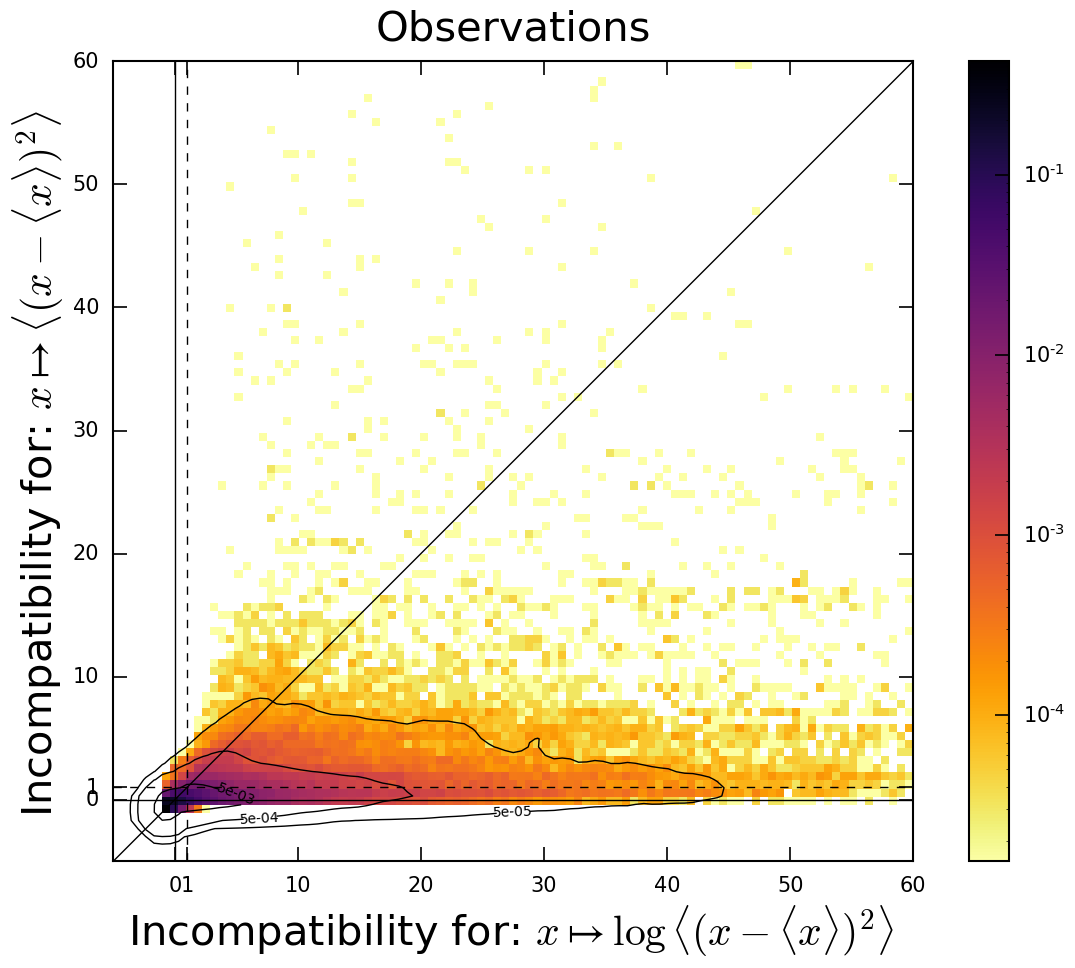}
    \caption{Effect of logarithmic rescaling after computing variance statistics on observation data. Provided that the assumptions of the test are not too discarded in both cases for these results to be appropriately interpreted on both axes, this shows that $\big\langle \big[x-\langle x \rangle_{\vec u}\big]^2 \big\rangle_{\vec u}$ is strictly more degenerate than $\log \, \big\langle \big[x-\langle x \rangle_{\vec u}\big]^2 \big\rangle_{\vec u}$.}
    \label{fig:log_better}
\end{figure}

Second, this could lead to dramatic improvement in incompatibility detection, as illustrated in Fig.~\ref{fig:log_better} for $\phi=\phi_{\text{var}}$ (provided that the assumptions of the test are not too discarded in both cases for these results to be appropriately interpreted).

\section{Apodization}\label{sec_apo}
\begin{figure}[htbp]
    \centering
    \includegraphics[width=.65\linewidth]{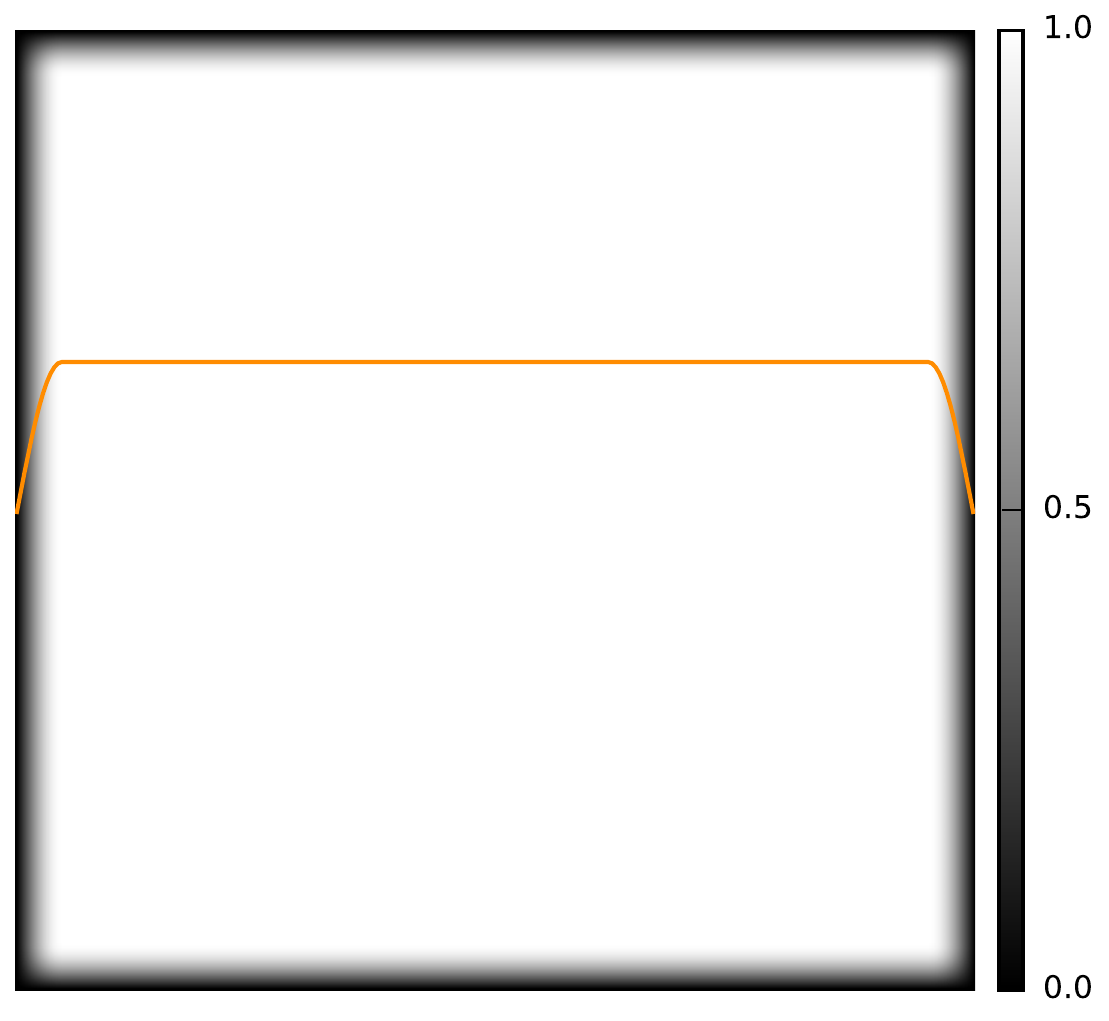}
    \caption{$256\times 256$ apodization window $w$ used. The orange line is a 1D slice of the middle of the window.}
    \label{fig:apo}
\end{figure}

The $256 \times 256$ subpatches on which we perform 2D Fourier transform do not have PBCs. Thus, before applying such transform during PS estimation on a given subpatch $x$, we first apodize it as follows:
$$x \mapsto r w \cdot (x-\langle x\rangle_{\vec u}),$$ where $w\cdot$ denotes the pixel-wise multiplication by the window $w$, reported in Fig.~\ref{fig:apo} and $r$ is a scalar factor depending on the input map $x$ such that the output apodized map has same variance as the input. 
The resulting map does not share the same mean with the input but this property is not regarded by the Fourier-based statistics used here.

\section{Additional examples of degeneracies}
\label{appendix:further_degeneracy_examples}

\begin{figure*}[htbp!]
\includegraphics[width=\linewidth]{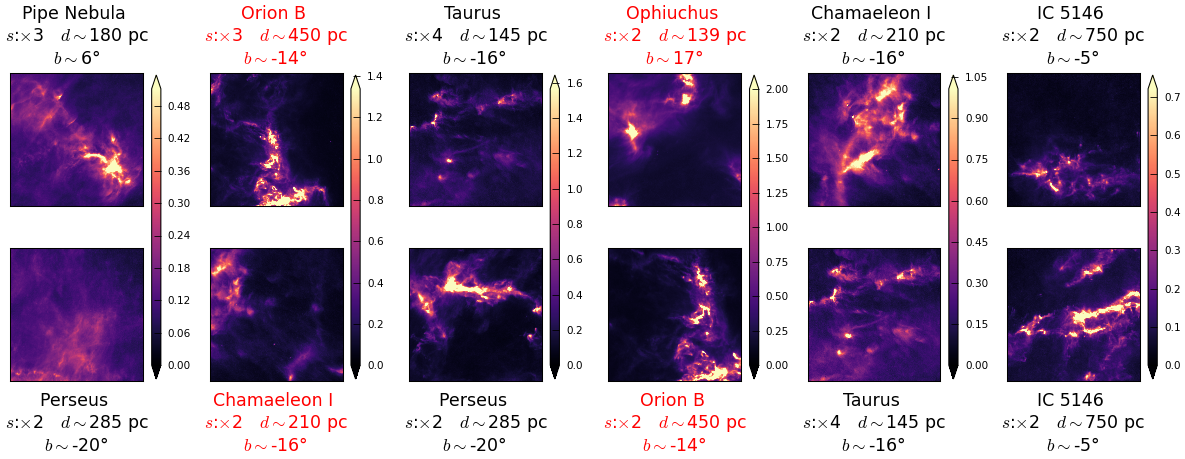}
\caption{Examples of log-Gaussian degeneracies. Six pairs of $512\times 512$ patches are chosen, whose locations on the scatter plot of Fig.~\ref{fig:loggaussStat_vs_RWST}.b are given there by the red stars. The column density maps are shown in units of $10^{20}\text{cm}^{-2}$. For each patch, we report: $s$ the subsampling factor from the original 3\arcsec/pix map, $d$ and $b$ the approximated distance and Galactic latitude of the cloud. If a pair has patches $(i,j)$ with incompatible pixel sizes according to the following criterion $\max \{\frac{s_id_i}{s_jd_j}, \frac{s_jd_j}{s_id_i} \} \geq 3/2$, we color its labels in red.}
\label{fig:loggaussStat_vs_RWST_OBS-1}
\end{figure*}

\begin{figure*}[htbp!]
\hspace*{-.5cm}\includegraphics[width=\linewidth]{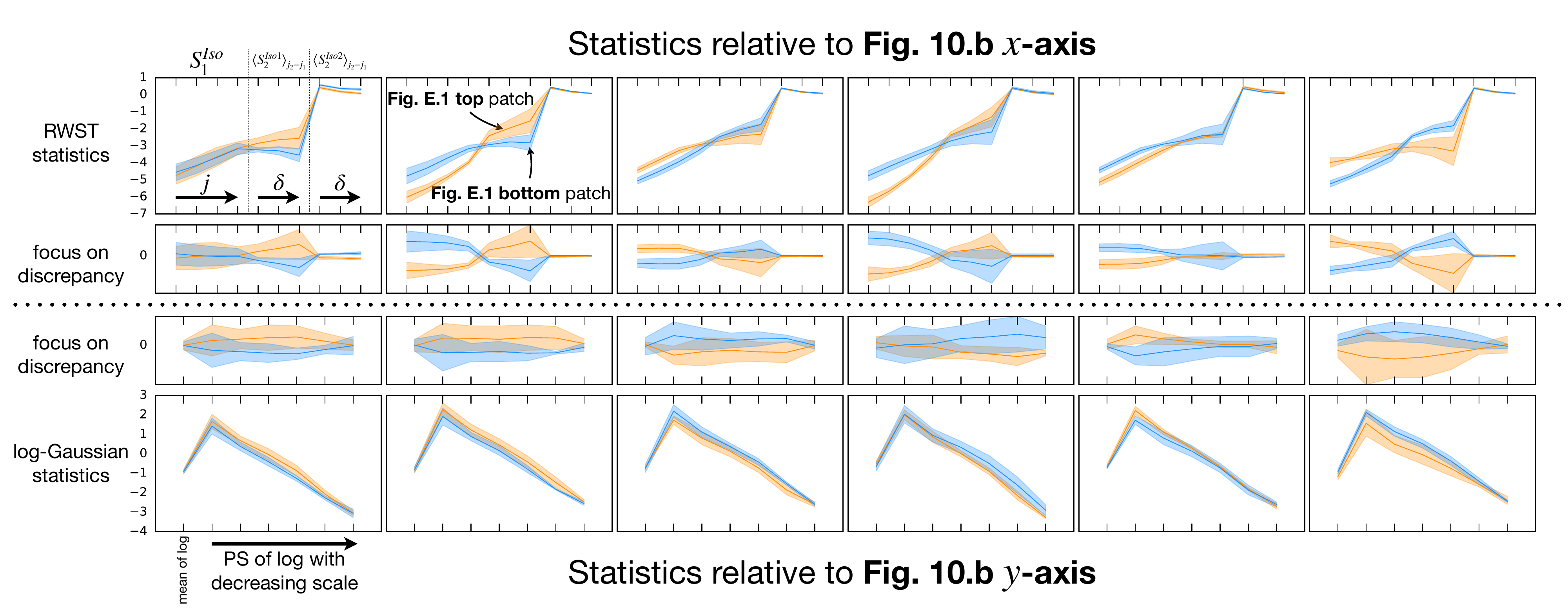}
\caption{Statistics for the examples of log-Gaussian confusions shown in Fig.~\ref{fig:loggaussStat_vs_RWST_OBS-1}. In each row, the orange filled line (resp. band) corresponds to the mean (resp. standard deviation) of the statistics computed over the four $256\times 256$ subpatches of the top patch of each pair of Fig.~\ref{fig:loggaussStat_vs_RWST_OBS-1}, and the corresponding blue lines and areas refer to the bottom patch of the pair. The top row corresponds to the RWST, starting with $S_1^{Iso}[j]$ coefficients with four increasing scales $j$, then $\langle S_2^{Iso1} \rangle_{j_2-j_1}[\delta]$ coefficients with three increasing scale ratios $\delta$ and finally $\langle S_2^{Iso2} \rangle_{j_2-j_1}[\delta]$ coefficients with the same three scale ratios. The bottom row represents in the following order: mean of log followed by PS of log with six decreasing scales. The second and third rows show the offsets of these statistics with respect to the mean of the two.}
\label{fig:loggaussStat_vs_RWST_OBS-2}
\end{figure*}

\EA{We show six examples of pairs of patches in Fig.~\ref{fig:loggaussStat_vs_RWST_OBS-1} that have compatible log-Gaussian statistics but incompatible RWST statistics, as shown in Fig.~\ref{fig:loggaussStat_vs_RWST_OBS-2}.}

\newpage
\section{Lower bound of the gap between observations and simulations}\label{appendix:mitigating_bias_distance}

\EA{In Sect.~\ref{SubSecIntObsSim}, we investigate whether observations and simulations do intersect. Ideally, we would like to check} whether the set of observations' moments $\{\bar \mu_i\}_{i \in \text{obs}}$ overlaps that of simulations $\{ \bar \mu_j \}_{j \in \text{sim}}$, where $\bar \mu_i \equiv \mathbb E [\phi_\text{final}(x_i)]$ stands for the expected value over a given process $i$. In practice, we would aim at retrieving the minimal distance\begin{equation}\bar d^2_{ij} \equiv (\bar \mu_i-\bar \mu_j)^T(\operatorname{diag} M_\text{obs})^{-1} (\bar \mu_i-\bar \mu_j)\end{equation}between $\bar \mu_i$ and $\bar \mu_j$ over (OBS, SIM) pairs $(i,j)$. 

However, the distance $d^2_\text{obs}$ introduced previously is a statistical estimator so that, in general,
\begin{equation*}
d^2_{ij} \neq   \bar d^2_{ij} ,
\end{equation*}
and in particular: $\mathbb E [d^2_{ij}] > \bar d^2_{ij} $. Indeed, the variance of the \EA{estimator} $\hat \mu_i$ biases $d^2_{ij}$ with respect to $\bar d^2_{ij}$:
\begin{equation}
\mathbb E [d^2_{ij}] \equiv \bar d^2_{ij} + b_{ij}.
\end{equation}
This non-negative bias, which boils down to \begin{equation} \label{eq_bias} b_{ij}=\mathrm{tr}\{\operatorname{cov} \big[ (\operatorname{diag} M_{\text{obs}})^{-1/2}(\hat \mu_i-\hat \mu_j) \big]\},\end{equation}prevents us from interpreting directly the value of $d^2_{ij}$ as $\bar d^2_{ij}$. Furthermore, its value increases with the amplitude of the fluctuations of $\hat \mu_i - \hat \mu_j$ and can thus change depending on the pair $(i,j)$ considered. This dependency on the pair also prevents us, without further check, from actually comparing $\bar d^2$ between different pairs based on the estimations $d^2$.

To probe and interpret the minimal distance between observations and simulations, we propose the following strategy:
\begin{itemize}
    \item Find an observation $i$ and a simulation $j$ that minimize the biased estimation $ d^2_{ij}$. Such pairs are already reported in Fig.~\ref{fig:best_match_globalVar_loggaussStat_and_RWST} and are good candidates to minimize $ \bar d^2_{ij}$. The goal then becomes to compare $ \bar d^2_{ij}$ with the minimal value $\bar d^2_{ii'}$, for $i'$ another observation, independent of $i$. But $\bar d^2$ is unknown. Instead:
    \vspace{0.1cm}
    \item Find an observation $i'$, that minimizes the biased distance $d^2_{ii'}$, while checking it does not overlap $i$ in the sky to assume them independent.
    \vspace{0.1cm}
    \item Since $\bar d^2_{ij} - \bar d^2_{ii'} = \mathbb E[d^2_{ij}] - \mathbb E[d^2_{ii'}] + b_{ii'}-b_{ij}$, the discrepancy between $\bar d^2_{ij} - \bar d^2_{ii'}$ can be estimated based on the measured discrepancy $ d^2_{ij} - d^2_{ii'}$ up to the unknown bias shift $b_{ii'}-b_{ij}$.
    \vspace{0.1cm}
    \item If, in addition the observation patch $x_{i'}$ is such that spatial fluctuations of $\phi_\text{final}$ over its subpatches are at least of the order of those estimated from the subpatches of the simulation patch $x_j$, then based on Eq.~\ref{eq_bias}, we have $b_{ii'}\gtrapprox b_{ij}$. This allows us then to use $ d^2_{ij} - d^2_{ii'}$ as an estimated lower bound for $ \bar d^2_{ij} -\bar d^2_{ii'}$.
\end{itemize}

We report in Fig.~\ref{fig:comparing_close_obs_sim} such independent\footnote{Indeed, only one of these pairs is made of patches retrieved from the same MC (Ophiuchus), but these patches do not overlap.} pairs $(i,i')$. In the first row, we fix $i$ associated with the patch from Ophiuchus $x_i$ that is the closest observation to simulations ($d^2_{ij}= 7$, Fig.~\ref{fig:best_match_globalVar_loggaussStat_and_RWST}). For this choice of patch, the reported distances $d^2_{ii'}$ of neighboring observations $i'$ are rather similar to $d^2_{ij}$. However, in the second row, where $i$ is associated with the patch from Aquila that is the second closest observation to simulations (as shown in Fig.~\ref{fig:best_match_globalVar_loggaussStat_and_RWST}), we see other observations $i'$, such as Serpens or Orion~B, that are closer than the closest simulation: $d^2_{ii'}= 3 < d^2_{ij}= 8$. We checked that the fluctuations of $i'$ are comparable to those of $j$ so that $b_{ii'}\gtrapprox b_{ij}$. This implies then $\bar d^2_{ij}-\bar d^2_{ii'} \gtrapprox 5 $ while $\bar d^2_{ii'} \leq d^2_{ii'} = 3$. Hence, $\bar d^2_{ij}/\bar d^2_{ii'} \gtrapprox 8/3 $. This second case shows that one of the closest (OBS, SIM) pairs is definitely further apart than this specific observation is with other observations, at least according to $\phi_\text{final}$. This example also supports that the potential bias on the $d^2_\text{obs}$ estimators remains moderate for the study of (OBS, SIM) pairs ($\bar d^2_{ij}\gtrapprox 5$ and $d^2_{ij}=8$ implies $b_{ij} \lessapprox 3$).

\end{appendix}

\end{document}